\documentclass[%
 reprint,
 amsmath,amssymb,
 aps,pre
]{revtex4-2}

\usepackage{graphicx}% Include figure files
\usepackage{dcolumn}% Align table columns on decimal point
\usepackage{bm}% bold math
\usepackage{hyperref}% add hypertext capabilities
\usepackage[utf8]{inputenc}

\usepackage{upgreek}

\usepackage{color}

\newcommand{\st}{\text{s}}
\newcommand{\hcs}{\text{HCS}}

\newcommand{\diag}{\text{dia}}
\newcommand{\ndiag}{\text{ndia}}

\newcommand{\mean}[1]{\left\langle #1 \right\rangle}

\DeclareMathOperator{\Tr}{Tr}
\DeclareMathOperator{\sgn}{sgn}

\begin{document}

\title{Linear response in the uniformly heated granular gas}
% Force line breaks with \\
%\thanks{A footnote to the article title}%

\author{Bernardo Sánchez-Rey}
\affiliation{Departamento de Física Aplicada I, E.P.S., Universidad de Sevilla, Virgen de África 7, E-41011 Sevilla, Spain}
\author{Antonio Prados}
\affiliation{Física Teórica, Universidad de Sevilla, Apartado de
  Correos 1065, E-41080 Sevilla, Spain}

\date{\today}% It is always \today, today,
             %  but any date may be explicitly specified

\begin{abstract}
  We analyse the linear response properties of the uniformly heated
  granular gas. The intensity of the stochastic driving fixes the
  value of the granular temperature in the non-equilibrium steady
  state reached by the system. Here, we investigate two specific
  situations. First, we look into the ``direct'' relaxation of the
  system after a single (small) jump of the driving intensity. This
  study is carried out by two different methods. Not only do we
  linearise the evolution equations around the steady state, but also
  derive generalised out-of-equilibrium fluctuation-dissipation
  relations for the relevant response functions. Second, we
  investigate the behaviour of the system in a more complex
  experiment, specifically a Kovacs-like protocol with two jumps in
  the driving. The emergence of anomalous Kovacs response is explained
  in terms of the properties of the direct relaxation function: it is
  the second mode changing sign at the critical value of the
  inelasticity that demarcates anomalous from normal behaviour. The
  analytical results are compared with numerical simulations of the
  kinetic equation, and a good agreement is found.
  %
% An article usually includes an abstract, a concise summary of the work
% covered at length in the main body of the article. 
% \begin{description}
% \item[Usage]
% Secondary publications and information retrieval purposes.
% \item[Structure]
% You may use the \texttt{description} environment to structure your abstract;
% use the optional argument of the \verb+\item+ command to give the category of each item. 
% \end{description}
\end{abstract}

%\keywords{Suggested keywords}%Use showkeys class option if keyword
                              %display desired
\maketitle

%\tableofcontents

\section{Introduction}\label{sec:intro}

Linear response is at the root of many important results in
physics. In this respect, the fluctuation-dissipation theorem is a
milestone in the development of (non-equilibrium) statistical
physics~\cite{kubo_fluctuation-dissipation_1966,callen_thermodynamics_1985,van_kampen_stochastic_1992}.
In its original form, it relates the linear relaxation of a system to
equilibrium, from an initial non-equilibrium state, with certain
equilibrium time correlation functions. Very recently, this result has
been extended to more general situations, which include the relaxation
to non-equilibrium steady states
(NESS)~\cite{marconi_fluctuationdissipation:_2008}.

The above general picture makes it relevant to investigate the linear
relaxation of intrinsically out-of-equilibrium systems, such as
granular
gases~\cite{goldhirsch_clustering_1993,brey_dissipative_1997,brilliantov_kinetic_2004,puglisi_fluctuations_2005},
to their NESS. Due to the energy dissipation in collisions, an
external energy input mechanism is needed to drive the system to a
NESS. One of the simplest physical situations is that of the uniformly
heated granular gas, in which all the particles of the gas are
submitted to independent white noise forces of a given amplitude
~\cite{van_noije_velocity_1998,montanero_computer_2000}. Therein, the
granular gas remains homogeneous for all times, if it was initially
so, and the main physical property of the granular gas is the granular
temperature---basically, the average kinetic energy per degree of
freedom. It is the amplitude of the white noise force, i.e. the
intensity of this \textit{stochastic thermostat}, that determines the
stationary value of the granular temperature.

The main purpose of this work is to analyse the linear relaxation of
the granular temperature in the uniformly heated granular gas, in
several different physical situations. Due to the non-Gaussian
character of the velocity distribution function, the velocity moments
obey an infinite hierarchy of coupled ordinary differential equations
(ODE), which must be supplemented with a closure assumption. Here, we
work in the first Sonine
approximation~\cite{goldshtein_mechanics_1995}, in which the state of
the granular system is characterised by the granular temperature $T$
and the second Sonine coefficient $a_{2}$---the excess kurtosis. The
dynamics of the gas is described by a set of two ODEs for the
temperature and the excess
kurtosis~\cite{garcia_de_soria_universal_2012}, first derived in
Ref.~\cite{huthmann_dynamics_2000} for the free cooling case. These
evolution equations are non-linear and have been shown to describe
accurately the granular gas in many different physical situations, see
for
example~\cite{van_noije_velocity_1998,montanero_computer_2000,brilliantov_deviation_2000,
  santos_second_2009,garcia_de_soria_universal_2012,prados_kovacs-like_2014,lasanta_when_2017}.

First, we would like to investigate the response of the gas to an
instantaneous perturbation of the driving intensity. This is a
relevant problem: different memory effects have been recently reported
in the uniformly heated granular
gas~\cite{prados_kovacs-like_2014,trizac_memory_2014,lasanta_when_2017}. In
principle, the existence of these memory effect suggest that the
linear relaxation of the granular temperature is
non-exponential. Indeed, non-exponential
relaxation~\cite{scherer_relaxation_1986,spohn_stretched_1989,kob_dynamics_1990,scherer_theories_1990,brey_stretched_1993,brey_low-temperature_1996,angell_relaxation_2000}
is a key ingredient for the emergence of aging and memory
effects~\cite{bouchaud_weak_1992,cugliandolo_evidence_1994,prados_aging_1997,josserand_memory_2000,brey_scaling_2007,ahmad_velocity_2007,hecht_aging_2017,lahini_nonmonotonic_2017,dillavou_nonmonotonic_2018,van_bruggen_tailoring_2019,keim_memory_2019,lulli_spatial_2020}
in many different physical contexts.

In light of the above, it is worth elucidating the relaxation
behaviour of the granular temperature in this single-jump experiment.
More specifically, we would like to clarify its exponential or
non-exponential character, and also its monotonicity properties. In
this respect, it is important to clarify the role played by the 
inelasticity. For example, a monotonic decay of the direct relaxation
function ensures that the Kovacs effect is ``normal'', i.e. the hump
has the same sign as in molecular systems~\cite{plata_kovacs-like_2017}. 

For small enough
perturbations, the evolution equations for $T$ and $a_{2}$ are
linearised and analytically solved. Moreover, the relaxation of both
quantities to their steady values is shown to be directly related to
some time correlation functions (calculated in the NESS), by extending
the ideas in Ref.~\cite{marconi_fluctuationdissipation:_2008} for an
out-of-equilibrium fluctuation-dissipation relation (FDR). The
correlation functions involve the derivative of the $N$-particle
velocity distribution function and thus we introduce a factorisation
assumption---sometimes called \textit{propagation of
  chaos}~\cite{garcia_de_soria_towards_2015}---to get an explicit
expression for the relevant time correlations.

Our analytical predictions are compared with Direct Simulation Monte
Carlo (DSMC) results---i.e. the numerical integration of the kinetic
equation. This is done for the two procedures described above. In the
``direct'' route, the system is initially put in the NESS
corresponding to a certain value $\xi_{0}$ of the driving, the driving
is instantaneously changed to $\xi$ at $t=0^{+}$, and subsequently the
time evolution of the granular temperature is recorded. In the FDR
route, the system is initially put in the NESS corresponding to $\xi$
and a the corresponding time correlation function is evaluated in this
NESS---there is no need to change the driving.

More detailed insight into the dynamics of the system can be acquired
with more complicated driving protocols. A particularly relevant one
is the so-called Kovacs experiment, which has been extensively
analysed in the realm of glassy
systems~\cite{kovacs_transition_1963,kovacs_isobaric_1979,buhot_kovacs_2003,bertin_kovacs_2003,arenzon_kovacs_2004,mossa_crossover_2004,aquino_kovacs_2006,prados_kovacs_2010,diezemann_memory_2011,di_signatures_2011,ruiz-garcia_kovacs_2014,banik_isochoric_2018,song_activation_2020}. The
Kovacs protocol involves two jumps of the parameter
$\xi$---temperature, driving intensity, etc.---controlling the
relaxation of the system. First, it is changed from $\xi_{0}$ to
$\xi_{1}<\xi_{0}$ at $t=0^{+}$ and the system relaxes towards the
steady---either equilibrium or NESS---state for $\xi_{1}$. This
relaxation lasts for a waiting time $t_{w}$: it is interrupted when
the value of the ``thermodynamic'' property of interest equals the
steady state value for some intermediate value of $\xi$, at
$t=t_{w}^{+}$ the value of the control is changed to $\xi$ ($\xi_{1}<\xi<\xi_{0}$). If the
subsequent behaviour is non-monotonic, with the thermodynamic property
departing from its steady value before returning thereto, additional
variables are needed to completely characterise the steady state of
the system.

On the one hand, the vast majority of the studies of the Kovacs effect
have been done in the non-linear regime, i.e. for large jumps of the
driving, and thus the results are mainly numerical. On the other hand,
a general theory of the Kovacs effect is only available within the
limits of applicability of linear response theory. For molecular
systems---such that the steady distribution is the canonical
one---with Markovian dynamics, it has been shown that he Kovacs hump
has some general characteristic features that stem from the standard
version of the FDR~\cite{prados_kovacs_2010}. Specifically, the Kovacs
hump has only one extremum and cannot change sign, it is always
\textit{normal}.

In athermal systems, there appear Kovacs responses that deviate from
the normal behaviour just described.  The Kovacs effect is said to be
\textit{anomalous} when only one extremum is present but the sign of
the hump changes with the system parameters. This behaviour has been
observed both in the uniformly heated granular
gas~\cite{prados_kovacs-like_2014,trizac_memory_2014} and in active
matter models~\cite{kursten_giant_2017}. Also, more than one extremum
has been reported in the rough granular
gas~\cite{lasanta_emergence_2019}. It must be stressed that all these
observations correspond to the non-linear regime. In particular, in
the uniformly heated granular case the extreme case $\xi_{1}=0$
(i.e. no driving in the waiting time window) was
considered~\cite{prados_kovacs-like_2014,trizac_memory_2014}.

The emergence of these more complex Kovacs responses, including the
anomalous behaviour, is not well understood on a general
basis. Although the existence of anomalous behaviour is supported by
analytical calculations in specific systems, the general physical
reason behind is not known. One of the main objectives of this work is
to shed light on this point by considering the linear response limit
in the uniformly heated granular gas. To the best of our knowledge,
the connection between the behaviour of the direct relaxation function
and the Kovacs hump has not been analysed in the context of granular
gases. Here, we show that the anomalous behaviour survives in linear
response and thus we can explain its emergence in terms of the
behaviour of the one-jump relaxation function. This is done by
bringing to bear recent results on the generalisation of the
mathematical structure of the linear Kovacs hump to athermal
systems~\cite{kursten_giant_2017,plata_kovacs-like_2017}.

The organisation of the paper is as follows. In
Sec.~\ref{sec:evol-eqs}, we put forward the evolution equations for
the granular temperature and the excess kurtosis and carry out their
linearisation around the NESS in Sec.~\ref{sec:linearisation}. The
linear relaxation of the granular temperature to an instantaneous
change of the driving is considered in Sec.~\ref{sec:linear-relax},
for different values of the inelasticity. The generalised FDR for a
jump in the driving is the subject of
Sec.~\ref{sec:check-neq-FDR}. Specifically, it is derived in
Sec.~\ref{sec:derivation-neq-FDR}, and particularised for the relevant
response functions in Sec.~\ref{sec:response-functions-FDR}. Section~\ref{sec:numerical-sim-linear-response} presents
DSMC results for the relaxation of the granular temperature and the
excess kurtosis---or, alternatively, the fourth-moment of the
velocity. Next,
Sec.~\ref{sec:kovacs-exp} is devoted to the analysis of the Kovacs
experiment. Finally, conclusions and a brief discussion of
perspectives for future work are presented in
Sec.~\ref{sec:discussion}.

\section{Evolution equations}\label{sec:evol-eqs}

Let us analyse the dynamics of a granular gas of smooth hard
spheres~\cite{brilliantov_kinetic_2004}. This system comprises $N$ hard particles of mass $m$ and
diameter $\sigma$ in dimension $d$, which undergo inelastic collisions. In the binary
collisions, the tangential component of the relative velocity remains
unaltered while the normal component is reversed and shrunk by a
factor $\alpha$, the normal restitution coefficient,
$0\leq\alpha\leq 1$. Apart from the elastic limit $\alpha=1$, kinetic
energy is lost in every collision and the undriven system ``cools'', in
the sense that its granular temperature---basically the average of the
kinetic energy---decreases monotonically in time.

We consider the uniformly heated granular gas, i.e. the system
described above is also submitted to a random forcing that inputs
energy into the system. This is modelled as independent white noise
forces $\bm{F}_{i}^{(p)}(t)$ acting over each grain,
$\langle F_{i}^{(p)}(t)\rangle=0$ and
$\langle
F_{i}^{(p)}(t)F_{j}^{(q)}(t')\rangle=m^{2}\xi^{2}\delta_{ij}\delta_{pq}
\delta(t-t')$, $i,j=1,\ldots,d$, $p,q=1,\ldots,N$. As a consequence,
the system reaches a steady state in the long time limit, in which the
energy input by the thermostat cancels---in average---the energy loss
in collisions~\cite{van_noije_velocity_1998}.

At the kinetic level of description, the dynamical evolution of the
system is governed by the Boltzmann-Fokker-Planck equation for the
velocity distribution
function~\cite{van_noije_velocity_1998,montanero_computer_2000}. Being
the granular gas an intrinsically non-equilibrium system, its velocity
distribution function is non-Gaussian, even in the steady state. The
granular temperature $T(t)$ is defined as
\begin{equation}
  \label{eq:granular-temp-def}
  \frac{d}{2}T(t)\equiv \left\langle \frac{1}{2}mv^{2}(t)\right\rangle.
\end{equation}
To measure the departure from the Maxwellian distribution, the
simplest approach is to consider the excess kurtosis $a_{2}$,
\begin{equation}
  \label{eq:a2-def}
  a_{2}=\frac{d}{d+2}\frac{\langle v^{4}\rangle}{\langle
    v^{2}\rangle^{2}}-1,
\end{equation}
which vanishes for the Gaussian case. In kinetic theory,
working with the pair $(T,a_{2})$ is known as the
first Sonine approximation, because $a_{2}$ is the first non-zero
coefficient of the expansion of the velocity distribution function in
a series of Sonine
polynomials~\cite{goldshtein_mechanics_1995,van_noije_velocity_1998,
  santos_second_2009,garzo_granular_2019},
\begin{equation}\label{eq:P1-Sonine}
  P_{1}(\bm{v};t)=v_{T}(t)^{-d}
  \pi^{-d/2}e^{-w^{2}}\left[1+a_{2}(t) 
    S_{2}\left(w^{2}\right)\right],
\end{equation}
where $\bm{w}$ is defined by
\begin{equation}
  \bm{w}=\frac{\bm{v}}{v_{T}(t)}, \qquad
  v_{T}^{2}(t)=\frac{2T(t)}{m}=\frac{2}{d}\langle v^{2}(t)\rangle,
\end{equation}
and
\begin{equation}
  S_{2}(x)=\frac{1}{2}x^{2}-\frac{d+2}{2}x+\frac{d(d+2)}{8}.
\end{equation}

The evolution equations for the granular temperature and the excess
kurtosis are derived from the Boltzmann-Fokker-Planck
equation~\cite{van_noije_velocity_1998,huthmann_dynamics_2000,santos_second_2009,garcia_de_soria_universal_2012,prados_kovacs-like_2014}. They
are usually written in terms of the collision rate $\zeta_{0}$, and the stationary values of the granular
temperature $T_{\st}$ and the excess kurtosis $a_{2}^{\st}$,
\begin{subequations}\label{eq:steady-values}
  \begin{equation}
    T_\st=\left[\frac{m\xi^2}
{\zeta_0(1+\frac{3}{16}a_2^\st)}\right]^{2/3}, \quad \zeta_0=\frac{2 n
\sigma^{d-1} \left(1-\alpha^2\right) \pi^{\frac{d-1}{2}}}{\sqrt{m}
d\Gamma(d/2)},
\end{equation}
\begin{equation}
  a_2^\st=\frac{16(1-\alpha)(1-2\alpha^2)}
{73+56d-24d\alpha-105\alpha+30(1-\alpha)\alpha^2}.
\end{equation}
\end{subequations}
By defining scaled---order of unity---variables as follows,
\begin{equation}\label{eq:scaled-variables-def}
\theta=\frac{T}{T_{\st}}, \quad
A_2=\frac{a_2}{a_2^\st}, \quad \uptau=\frac{\zeta_0 \sqrt{T_\st}}{2}
t.
\end{equation}
we get
\begin{subequations}\label{eq:T-and-a2-evol-eqs}
  \begin{equation}\label{eq:T-and-a2-evol-eqsa}
    \frac{d\theta}{d\uptau}=2\left[ 1-\theta^{3/2}
+\frac{3}{16}a_2^\st \left( 1-A_{2}\theta^{3/2} \right)\right],
\end{equation}
\begin{equation}\label{eq:T-and-a2-evol-eqsb}
  \theta
  \frac{dA_{2}}{d\uptau}=4 \left[\left(\theta^{3/2}-1\right) A_2+
    B\, \theta^{3/2} \left( 1-A_2 \right) \right],
\end{equation}
\end{subequations}
with the parameter $B$ given by~\footnote{This expression for $B$ stems from
  a lengthy kinetic theory
  calculation~\cite{garcia_de_soria_universal_2012} or, preferably,
  from a simple self-consistency
  argument~\cite{prados_kovacs-like_2014,trizac_memory_2014}.}
\begin{equation}\label{eq:B-param}
B =\frac{73+8d(7-3\alpha)+15\alpha[2\alpha(1-\alpha)-7]}
{16(1-\alpha)(3+2d+2\alpha^2)}.
\end{equation}
A couple of comments on the evolution
equations~\eqref{eq:T-and-a2-evol-eqs} are appropriate. First, they
are non-linear, so in general it is not possible to write down an
analytical closed solution for them~\footnote{They are non-linear in
  the temperature but linear in the excess kurtosis. This is a
  consequence of the first Sonine approximation, in which
  non-linearities in $a_{2}$ are neglected.}. Second, the parameters
$B$ and $a_{2}^{\st}$ only depend on $\alpha$ and $d$, being
independent of the steady value of the temperature. In fact, all the
dependence on $T_{\st}$ has been incorporated into the time scale
$\uptau$ introduced in Eq.~(\ref{eq:scaled-variables-def}).

It will be useful for our purposes to define the dimensionless
velocity
\begin{equation}\label{eq:c-def}
  \bm{c}=\frac{\bm{v}}{v_{T}^{\st}}, \quad
  v_{T}^{\st}=\sqrt{\frac{2T_{\st}}{m}},
\end{equation}
the second moment of which is directly related to the dimensionless temperature,
\begin{equation}
  \mean{c^{2}}=\frac{\mean{v^{2}}}{\left(v_{T}^{\st}\right)^{2}}=\frac{d}{2}\theta,
  \qquad \mean{c^{2}}_{\st}=\frac{d}{2}.
\end{equation}
Equation~(\ref{eq:a2-def}) still holds with the change $v\to c$,
whereas the fourth moment of $c$ at the steady state is given by
\begin{equation}\label{eq:c4-av-st}
  \langle c^{4}\rangle_{\st}=\frac{d+2}{d}\langle
  c^{2}\rangle_{\st}^{2}(1+a_{2}^{\st})=\frac{d(d+2)}{4}(1+a_{2}^{\st}). 
\end{equation}

\subsection{Linearisation around the steady
  state}\label{sec:linearisation}

In this paper, we are interested in the linear relaxation of the
granular gas to the steady state, in which the temperature and the
excess kurtosis are $\theta_{\st}=1$ and $A_{2}^{\st}=1$, as a
consequence of the scaling. Therefore, we write
\begin{equation}\label{eq:delta-theta-A2-def}
 \theta=1+\delta\theta, \qquad A_{2}=1+\delta A_{2}, 
\end{equation}
insert them into Eq.~\eqref{eq:T-and-a2-evol-eqs} and neglect
nonlinearities. The following linear system is thus obtained,
\begin{subequations}\label{eq:delta-theta-a2-evol-eq}
  \begin{equation}
    \frac{d}{dt}\delta\bm{z}=\bm{M}\cdot\delta\bm{z},
  \end{equation}
  \begin{equation}
    \delta\bm{z}\equiv\begin{pmatrix}
    \delta\theta \\ \delta A_{2} \end{pmatrix}, \qquad
  \bm{M}\equiv\begin{pmatrix} -3\left(1+\frac{3}{16}a_{2}^{\st}\right) &
    -\frac{3}{8}a_{2}^{\st} \\ 6 & -4B \end{pmatrix}.
  \end{equation}
\end{subequations}
The matrix---or operator~\footnote{We are employing a notation similar to that introduced in
Ref.~\cite{plata_kovacs-like_2017} to investigate the linear response
of a general athermal system, at the level of description of the
equations for the moments.}---$\bm{M}$ does not depend on the noise
intensity $\xi$ but only on the restitution coefficient $\alpha$ and
the dimension $d$. This is a consequence of the rhs of the non-linear
system \eqref{eq:T-and-a2-evol-eqs} being independent of $T_{\st}$, as
already pointed out above.

In the following, we denote the elements of the matrix $\bm{M}$ by
$M_{ij}$, i.e.
\begin{align}
  \label{eq:Mij-def} M_{11}&=-3\left(1+\frac{3}{16}a_{2}^{\st}\right),
& M_{12}&=-\frac{3}{8}a_{2}^{\st}, \nonumber \\ \ M_{21}&=6, &
M_{22}&=-4B.
\end{align}
The eigenvalues of the matrix $\bm{M}$ are both negative,
\begin{equation}
  \label{eq:M-eigenval}
  \lambda_{\pm}=\frac{\Tr\bm{M}\pm\sqrt{(\Tr\bm{M})^{2}-
      4\det\bm{M}}}{2}<0,
\end{equation}
and the corresponding eigenvectors read
\begin{equation}
  \label{eq:M-eigenvec}
  \bm{\zeta}_{+}=\begin{pmatrix} M_{12} \\
    \lambda_{+}-M_{11} \end{pmatrix}, \quad
  \bm{\zeta}_{-}=\begin{pmatrix} M_{12} \\
    \lambda_{-}-M_{11} \end{pmatrix}.
\end{equation}
The explicit expression for the trace and the determinant of $\bm{M}$
are
\begin{subequations}
\begin{align}
  \label{eq:trace-and-det-M}
  \Tr\bm{M}& =-4B-3\left(1+\frac{3}{16}a_{2}^{\st}\right)<0, \\
  \det\bm{M}&=12B\left(1+\frac{3}{16}a_{2}^{\st}\right)+
  \frac{9}{4}a_{2}^{\st}>0.
\end{align}
\end{subequations}

The general solution of the linear system is
\begin{equation}
  \label{eq;z-sol}
  \delta\bm{z}(\uptau)=C_{+}\bm{\zeta}_{+}e^{\lambda_{+}\uptau}+
  C_{-}\bm{\zeta}_{-}e^{\lambda_{-}\uptau},
\end{equation}
or, equivalently,
\begin{equation}
  \label{eq:z-sol-bis}
  \begin{pmatrix}
    \delta\theta(\uptau) \\ \delta\!A_{2}(\uptau)
  \end{pmatrix}\! =
  C_{+}\! \begin{pmatrix} M_{12} \\
    \lambda_{+}\!-M_{11} \end{pmatrix}
  e^{\lambda_{+}\uptau} +
  C_{-}\! \begin{pmatrix} M_{12} \\
    \lambda_{-}\!\!-M_{11} \end{pmatrix}
  e^{\lambda_{-}\!\uptau}.  
\end{equation}
The constants $C_{+}$ and $C_{-}$ are determined by the
initial conditions through the relation
\begin{equation}
  \label{eq:c+-and-c-}
  \begin{pmatrix}
    \delta\theta(0) \\ \delta\!A_{2}(0)
  \end{pmatrix}= \delta\bm{z}(0)=C_{+}\bm{\zeta}_{+}+C_{-}\bm{\zeta}_{-}.
\end{equation}

%%%%%%%%%%%%%%%%

\section{Direct relaxation after a single jump in the driving}\label{sec:linear-relax}

Let us consider an experiment in which the system is initially at the
steady state corresponding to some value of the driving
$\xi+\delta\xi$. At the initial time, the intensity of the driving is
suddenly changed from $\xi+\delta\xi$ to $\xi$: subsequently, the
granular gas relaxes from its non-equilibrium steady state (NESS) at
$\xi+\delta\xi$ to the corresponding NESS at $\xi$.

The initial values of the scaled variables are
\begin{equation}
  \label{eq:init-val-temp-relax}
  \delta\theta(0)\neq 0, \quad
  \delta\! A_{2}(0)=0,
\end{equation}
because the steady value of the temperature depends on the driving
intensity but that of the excess kurtosis does
not. Eq.~(\ref{eq:c+-and-c-}) predicts that both $C_{+}$
and $C_{-}$ are proportional to
$\delta\theta(0)$ in this case, so we write
\begin{equation}
  C_{\pm}=\delta\theta(0)\,\tilde{C}_{\pm}
\end{equation}
and readily obtain
\begin{equation}
  \label{eq:c+-c--temp-relax}
  \tilde{C}_{+}=\frac{M_{11}-\lambda_{-}}
  {M_{12}\left(\lambda_{+}-\lambda_{-}\right)}, \quad
  \tilde{C}_{-}=\frac{\lambda_{+}-M_{11}}
  {M_{12}\left(\lambda_{+}-\lambda_{-}\right)}.
\end{equation}
Therefore, we have that
\begin{equation}
  \label{eq:phi1}
  \frac{\delta\bm{z}(\uptau)}{\delta\theta(0)}=
  \tilde{C}_{+}\bm{\zeta}_{+}e^{\lambda_{+}\uptau}+
  \tilde{C}_{-}\bm{\zeta}_{-}e^{\lambda_{-}\uptau}.
\end{equation}

In what follows, we denote the relaxation function for the physical
property $Y$ by $\phi_{Y}(\uptau)$, specifically
\begin{equation}
  \phi_{Y}(\uptau)=\frac{\delta Y(\uptau)}{\delta\theta(0)}
\end{equation}
First, we analyse the relaxation of the granular temperature, i.e.
\begin{equation}
  \label{eq:phi11}
 \phi_{\theta}(\uptau)\equiv\frac{\delta\theta(\uptau)}{\delta\theta(0)}=
  a_{+}e^{\lambda_{+}\uptau}+a_{-}e^{\lambda_{-}\uptau},
\end{equation}
with
\begin{equation}
  \label{eq:a+-and-a-}
  a_{+}=\frac{M_{11}-\lambda_{-}}{\lambda_{+}-\lambda_{-}}, \quad
  a_{-}=\frac{\lambda_{+}-M_{11}}{\lambda_{+}-\lambda_{-}}.
\end{equation}

\begin{figure}
  \centering
  \begin{tabular}{r}
    \includegraphics[width=2.7in]{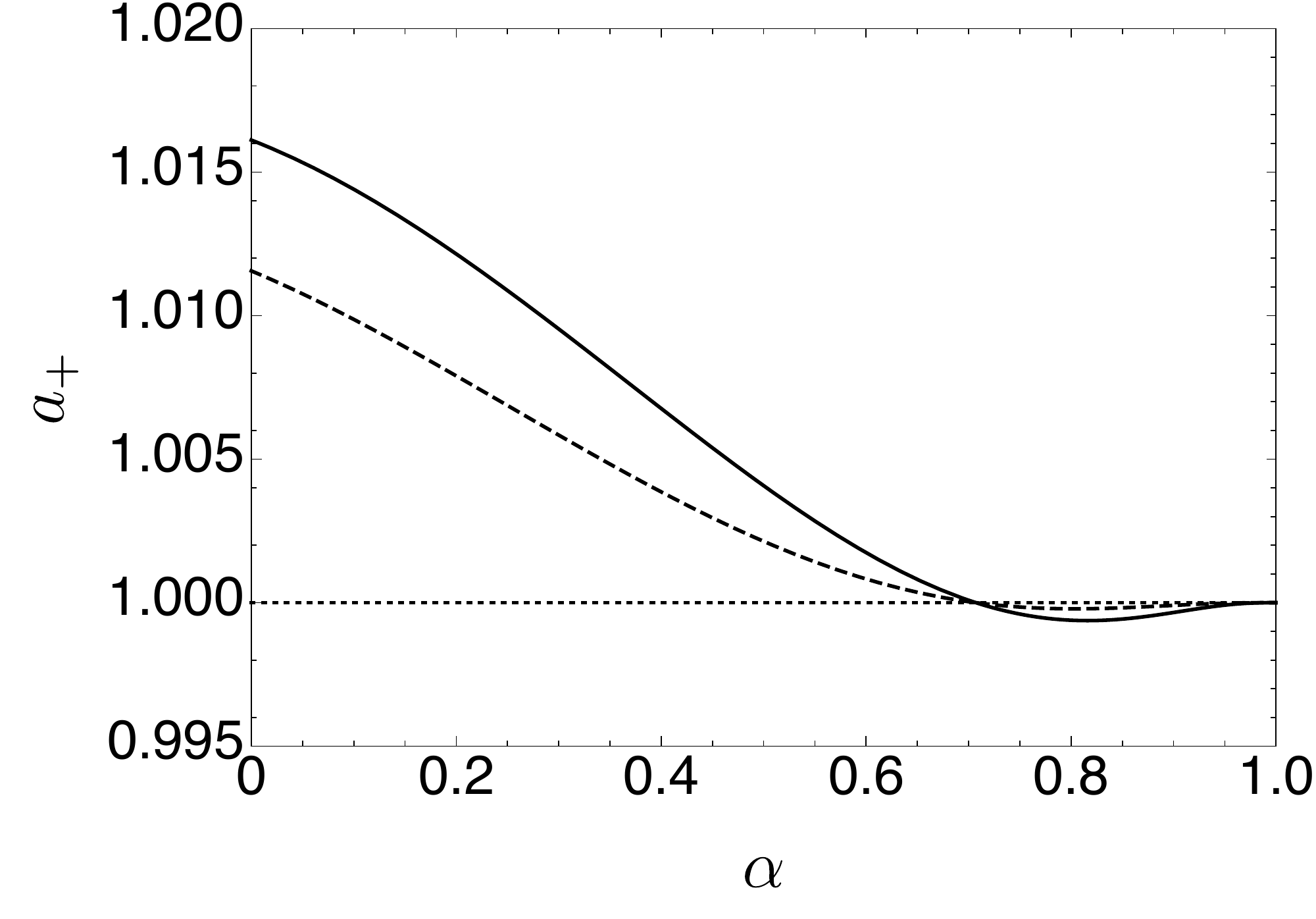}\\
    \includegraphics[width=2.75in]{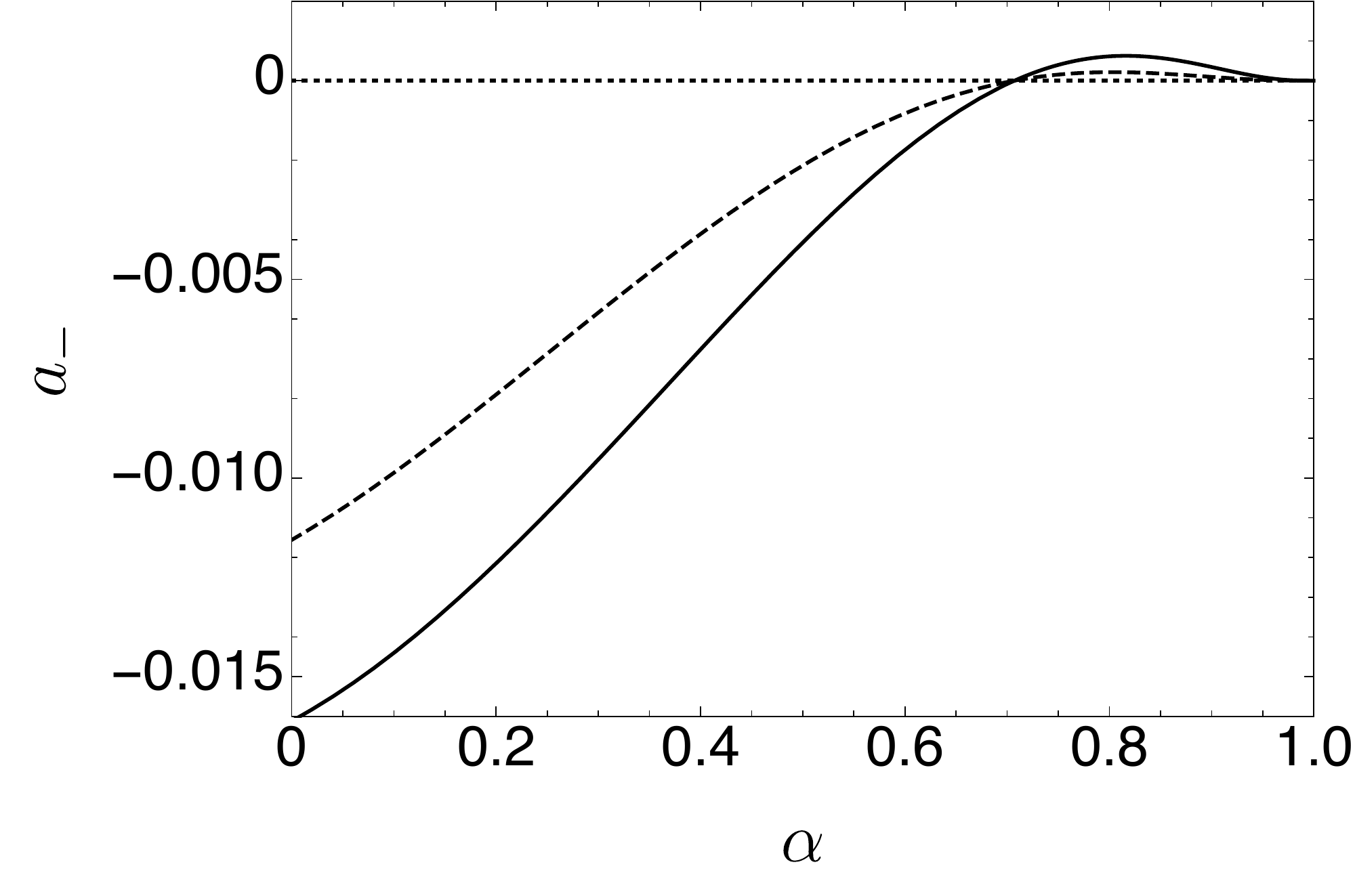}
  \end{tabular}
  \caption{\label{fig:a+-and-a-}Coefficients $a_{\pm}$ as a function
    of the restitution coefficient $\alpha$. Both the
    three-dimensional (solid line) and the two-dimensional (dashed
    line) are plotted. The coefficient $a_{+}$ is always positive,
    whereas $a_{-}$ changes sign at the critical value
    $\alpha=\alpha_{c}$---as given by Eq.~\eqref{eq:sgn-a2}. Note that
    $|a_{-}|$ is quite small throughout, with its least small value at
    $\alpha=0$, and thus $a_{+}=1-a_{-}$ is always very close to
    unity. }
\end{figure}

In Fig.~\ref{fig:a+-and-a-}, we show the two coefficients $a_{+}$ and
$a_{-}$ as a function of the restitution coefficient $\alpha$, for
$d=2$ and $d=3$. Both for the two and the three-dimensional case, two
distinct properties are observed: (i) while $a_{+}>0$ for all values
of $\alpha$, $a_{-}$ changes sign at the critical value
$\alpha_{c}=1/\sqrt{2}$, specifically
\begin{equation}
  \label{eq:sgn-a2}
  \sgn(a_{-})=\sgn(\alpha-\alpha_{c}), \quad \alpha_{c}=1/\sqrt{2},
\end{equation}
and (ii) the ratio $|a_{-}/a_{+}|\ll 1$ for all values
of $\alpha$, with its larger value  taking
place at $\alpha=0$, for which it equals $0.0159$ ($0.0114$) for $d=2$ ($d=3$). In principle,
the change of sign of $a_{-}$ at $\alpha_{c}$ might bring about a
non-monotonic behaviour of the relaxation function, but this is not
the case. In the long time limit, the relaxation is dominated by the
largest eigenvalue $\lambda_{+}$ and, therefore,
\begin{equation}
  \label{eq:phi1-long-times}
  \phi_{\theta}(\uptau)\sim a_{+}e^{\lambda_{+}\uptau}>0, \quad
  (\lambda_{+}-\lambda_{-})\uptau\gg 1.
\end{equation}

In order to understand the above behaviour of $a_{\pm}$, we bring to
bear the smallness of $a_{2}^{\st}$ and expand all quantities in
powers of $a_{2}^{\st}$. In particular, the eigenvalues
$\lambda_{\pm} $ are given by
\begin{align}
  \lambda_{+}&=-3-\frac{9(1+4B)}{16(4B-3)}a_{2}^{\st}+O((a_{2}^{\st})^{2}),
               \label{eq:lambda+-expansion}
  \\
  \lambda_{-}&=-4B+\frac{9}{4(4B-3)}a_{2}^{\st}+O((a_{2}^{\st})^{2})
               \label{eq:lambda--expansion}.
\end{align}
Note that $4B-3>0$ ($\lambda_{+}>\lambda_{-}$). In turn, the
corresponding expansion for the coefficients $a_{\pm}$ is
\begin{align}
  a_{+}&=1+\frac{9}{4(4B-3)^{2}}a_{2}^{\st}+O((a_{2}^{\st})^{2}),
         \label{eq:a+-expansion}
  \\
  a_{-}&=-\frac{9}{4(4B-3)^{2}}a_{2}^{\st}+O((a_{2}^{\st})^{2})
         \label{eq:a--expansion}.
\end{align}
On the one hand, it is neatly observed that $a_{-}$ is proportional to
$a_{2}^{\st}$, thus being rather small across the whole range of
inelasticities. Moreover, the change of sign of $a_{-}$ at
$\alpha_{c}$ is clearly linked to the vanishing of
$a_{2}^{\st}$ thereat. On the other hand, $a_{+}=1+O(a_{2}^{\st})$ is
very close to unity throughout, consistently with the normalisation
condition $a_{+}+a_{-}=1$.

Additional information is given by the relaxation of the excess
kurtosis or, equivalently, of the fourth moment $\langle
c^{4}\rangle$. Eq.~\eqref{eq:phi1} implies
\begin{equation}\label{eq:phi1-2}
  \phi_{\!A_{2}}(\uptau)\equiv
  \frac{\delta\!
    A_{2}(\uptau)}{\delta\theta(0)}=
  \frac{a_{+}a_{-}(\lambda_{+}-\lambda_{-})}
  {M_{12}}
  \left(e^{\lambda_{+}\uptau}-e^{\lambda_{-}\uptau}\right).
\end{equation}
The coefficient in front of the difference of exponentials can also be
expanded in powers of $a_{2}^{\st}$, making use of
Eqs.~(\ref{eq:Mij-def}) and
\eqref{eq:lambda+-expansion}--\eqref{eq:a--expansion}, with the result
\begin{equation}\label{eq:positiviness-A2final}
  \frac{a_{+}a_{-}(\lambda_{+}-\lambda_{-})}
  {M_{12}}=\frac{6}{4B-3}+\frac{27(5+4B)}{8(4B-3)^{3}}a_{2}^{\st}
  +O((a_{2}^{\st})^{2}).
\end{equation}
The dominant behaviour $6/(4B-3)$ is positive for all $\alpha$: this
means that $\delta A_{2}(\uptau)$ has the same sign that
$\delta\theta(0)$ for all times, i.e. the same sign of
$\delta\xi$~\footnote{The steady value of the granular
temperature is an increasing function of the driving, both
$\delta\theta(0)$ and $\delta\xi$ have the same sign.}. Note that, in our experiment, the driving is
instantaneously decreased from $\xi+\delta\xi$ to $\xi$ at the
initial time: therefore $\delta\xi>0$ and $\delta\!A_{2}(\uptau)>0$.

The relaxation function $\phi_{\langle c^{4}\rangle}(\tau)$ can be
written as a linear combination of $\phi_{\theta}$ and
$\phi_{\!A_{2}}$. Bringing to bear Eqs.~\eqref{eq:a2-def}, \eqref{eq:c-def}
, \eqref{eq:phi11}, and \eqref{eq:phi1-2}, one gets
\begin{equation}
  \delta\langle
  c^{4}\rangle=\frac{d(d+2)}{2}\left[\frac{1}{2}
    a_{2}^{\st}\,\delta\!
    A_{2}+(1+a_{2}^{\st})\delta\theta\right] 
\end{equation}
for the deviation of $\langle c^{4}\rangle$ from its steady state
value, given by Eq.~\eqref{eq:c4-av-st}. The relaxation function thus reads
\begin{equation}\label{eq:phic4-direct}
  \phi_{\langle c^{4}\rangle}(\uptau)=\frac{d(d+2)}{2}\left[\frac{1}{2}
    a_{2}^{\st}\,
    \phi_{\!A_{2}}(\uptau)+(1+a_{2}^{\st})\phi_{\theta}(\uptau)\right] .
\end{equation}

\section{Generalised FDR for a jump in the driving}\label{sec:check-neq-FDR}

The relaxation functions above can be related to certain time
correlations in the NESS of the granular gas, by means of a
generalised FDR. In the following,  we first derive a generalised FDR
for the relaxation of an arbitrary function of the velocities
and afterwards particularise it to get the specific relations for the
relaxation functions considered in the previous section.

\subsection{Derivation of the generalised
  FDR}\label{sec:derivation-neq-FDR}

Now we investigate the evolution of the granular gas, after an
instantaneous jump in the driving like the one considered in the
previous section. At the $N$-particle level, the fluid is described by
$P_{N}(\bm{\Gamma},t)$, which is the (one-time) probability density
for finding the system with
$\bm{\Gamma}=(\bm{v}_{1},\bm{v}_{2},\ldots,\bm{v}_{N})$ at time $t$.

For homogeneous situations---like those considered throughout
this paper, $\bm{\Gamma}$ is a Markov process described by Kac's
equation~\cite{kac_foundations_1956,garcia_de_soria_towards_2015} and
for any function of the velocities we can write
\begin{align}\label{eq:fGamma-mean}
 \mean{f(\bm{\Gamma}_t)}&=\int d\bm{\Gamma}_{t} f(\bm{\Gamma}_{t})
                           P_{N}(\bm{\Gamma}_{t},t) \nonumber \\
  &= \int d\bm{\Gamma}_{t}\int d\bm{\Gamma}_{0}
    f(\bm{\Gamma}_{t}) T_{t}(\bm{\Gamma}_{t}|\bm{\Gamma}_{0})
    P_{N}(\bm{\Gamma}_{0},0), 
\end{align}
where $T_{t}(\bm{\Gamma}_{t}|\bm{\Gamma}_{0})$ is the transition
probability from $\bm{\Gamma}_{0}$ to $\bm{\Gamma}_{t}$ in a time
interval $t$. Thus, we have
\begin{equation}
  P_{N}(\bm{\Gamma},t,\bm{\Gamma'},0)=
  T_{t}(\bm{\Gamma}|\bm{\Gamma'})P_{N}(\bm{\Gamma'},0)
\end{equation}
for the two-time probability density for finding the $N$-particle
granular gas with
$\bm{\Gamma}=(\bm{v}_{1},\bm{v}_{2},\ldots,\bm{v}_{N})$ at time $t$
and with $\bm{\Gamma'}=(\bm{v'}_{1},\bm{v'}_{2},\ldots,\bm{v'}_{N})$
at time $t=0$.

Now we consider that the initial distribution at $t=0$
corresponds to the steady state for $T(0)=T_{\st}+\delta T(0)$, 
\begin{equation}
  P_{N}(\bm{\Gamma}_{0},0)\!= \!
  P_{N}^{\st}(\bm{\Gamma}_{0})+\delta T(0)\,
  \frac{\partial P_{N}^{\st}(\bm{\Gamma}_{0})}{\partial T_{\st}}.
\end{equation}
Inserting this equation into Eq.~\eqref{eq:fGamma-mean}, one gets
\begin{equation}\label{eq:deltaf-Gamma-mean}
  \delta\mean{f(\bm{\Gamma}_{t})}=\delta T(0) \int\!\! d\bm{\Gamma}_{t}
  \int\!\! d\bm{\Gamma}_{0}
  f(\bm{\Gamma}_{t}) T(\bm{\Gamma}_{t}|\bm{\Gamma}_{0})
  \frac{\partial P_{N}^{\st}(\bm{\Gamma}_{0})}{\partial T_{\st}},
\end{equation}
in which we have defined, consistently with the notation we employ
throughout,
\begin{equation}
  \delta\mean{f(\bm{\Gamma}_{t})}\equiv
  \mean{f(\bm{\Gamma}_{t})}-\mean{f(\bm{\Gamma})}_{\st} 
\end{equation}
for the deviation of the average value of $f$ from its steady state
value.

Equation~\eqref{eq:deltaf-Gamma-mean} can be rewritten as
\begin{equation}\label{eq:generalised-FDR}
  \delta\mean{f(\bm{\Gamma}_{t})}=\delta T(0)
  \mean{f(\bm{\Gamma}_{t})\frac{\partial\ln\!
      P_{N}^{\st}(\bm{\Gamma}_{0})}{\partial T_{\st}}}_{\!\!\st} ,
\end{equation}
there emerges the time correlation function
\begin{align}
  &C_{\! f}(t)\equiv\mean{f(\bm{\Gamma}_{t})\frac{\partial\ln\!
  P_{N}^{\st}(\bm{\Gamma}_{0})}{\partial T_{\st}}}_{\!\!\st} \nonumber \\
  &\;\;= \int\!\! d\bm{\Gamma}_{t}
  \int\!\! d\bm{\Gamma}_{0} f(\bm{\Gamma}_{t})\frac{\partial\ln\!
    P_{N}^{\st}(\bm{\Gamma}_{0})}{\partial T_{\st}}
  T(\bm{\Gamma}_{t}|\bm{\Gamma}_{0}) P_{N}^{\st}(\bm{\Gamma}_{0}).
\end{align}
Therefore, we can also obtain the relaxation functions by measuring
correlations at the NESS, which is the expression of the generalised
FDR. The subindex $s$ in $\mean{\cdots}_{\st}$ stresses that the
average in the time correlation function is done in the steady state,
because
\begin{equation}
  T(\bm{\Gamma}_{t}|\bm{\Gamma}_{0}) P_{N}^{\st}(\bm{\Gamma}_{0})=
  P_{N}^{\st}(\bm{\Gamma}_{t},t,\bm{\Gamma}_{0},0)
\end{equation}
is the two-time probability density at the steady state with
temperature $T_{\st}$~\footnote{ The derivation above follow the same
  lines of the proof of a FDR in non-Hamiltonian systems of
  Ref.~\cite{marconi_fluctuationdissipation:_2008}. Therein, the
  authors consider the response to an initial perturbation of the form
  $\bm{\Gamma}_{0}\to \bm{\Gamma}_{0}+\delta\bm{\Gamma}_{0}$.}.

The difficulty of this approach is the necessity of knowing the
$N$-particle distribution $P_{N}^{\st}(\bm{\Gamma}_{0})$: note that
one deals with the one-particle distribution function in the kinetic
approach. This difficulty is usually circumvented by introducing the
following factorisation
ansatz~\cite{marconi_fluctuationdissipation:_2008}
\begin{equation}\label{eq:propag-chaos}
  P_{N}^{\st}(\bm{\Gamma})=\prod_{i=1}^{N} P_{1}(\bm{v}_{i}),
\end{equation}
which is sometimes called ``propagation of
chaos''~\cite{garcia_de_soria_towards_2015}. This is more restrictive
than the usual \textit{molecular chaos} hypothesis employed to derive
the Boltzmann equation: therein, only the two-particle distribution is
assumed to factorise. With the assumption~\eqref{eq:propag-chaos},
$C_{\! f}(t)$ simplifies to
\begin{align}
  C_{\!f}(t)=\sum_{i=1}^{N}C_{\! f}^{i}(t), \quad C_{\! f}^{i}(t)\equiv
  \mean{f(\bm{\Gamma}_{t})\frac{\partial\ln\!
  P_{1}^{\st}(\bm{v}_{i0})}{\partial T_{\st}}}_{\!\!\st} ,
\end{align}
where $P_{1}^{\st}$ is the stationary solution of the Boltzmann
equation, which is---at least approximately---known.

\subsection{Response functions}\label{sec:response-functions-FDR}

Here, we derive the specific formulas relating the relaxation
functions $\phi_{\theta}$ and $\phi_{\langle c^{4}\rangle}$ to time
correlations.  To achieve this goal, we need
$\partial_{T_{\st}}\ln\!P_{N}^{\st}(\bm{\Gamma})$ and thus we use the
propagation of chaos assumption~\eqref{eq:propag-chaos}. Consistently
with our approach, we employ the one-particle velocity distribution
function in the first Sonine approximation. Particularising
\eqref{eq:P1-Sonine} to the steady state, we have
\begin{equation}\label{eq:P1s-Sonine}
  P_{1}^{\st}(\bm{v})=\left(v_{T}^{\st}\right)^{-d}e^{-c^{2}}\left[1+a_{2}^{\st}
  S_{2}\left(c^{2}\right)\right],
\end{equation}
where $\bm{c}$ is defined in Eq.~\eqref{eq:c-def}. Taking into account
that $\partial c^{2}/\partial T_{\st}=-c^{2}/T_{\st}$, we get
\begin{equation}
  \frac{\partial\!
    \ln\!P_{1}(\bm{v})}{\partial T_{\st}}
  =-\frac{d}{2T_{\st}}+\frac{1}{T_{\st}}c^{2}-\frac{1}{T_{\st}}c^{2}  
  \frac{a_{2}^{\st}S_{2}^{\prime}
    \left(c^{2}\right)}{1+a_{2}^{\st}
  S_{2}\left(c^{2}\right)}.
\end{equation}
By defining
\begin{equation}
  g(x)=-\frac{d}{2}+x-x \frac{a_{2}^{\st}S_{2}'(x)}{1+a_{2}^{\st}S_{2}(x)},
\end{equation}
we can write
\begin{equation}
  \frac{\partial\!
    \ln\!P_{1}(\bm{v})}{\partial
    T_{\st}}=\frac{1}{T_{\st}}g\!\left(c^{2}\right), \quad
  C_{f}^{i}(t)=\frac{1}{T_{\st}}\mean{f(\bm{\Gamma}_{t})\,
    g\!\left(c_{i0}^{2}\right)}_{\st}.
\end{equation}
and thus the response function is given by 
\begin{equation}\label{eq:FDR-C-f-g-sum}
  \delta \mean{f(\bm{\Gamma}_{t})}=\delta\theta(0)\sum_{i=1}^{N}
  \mean{f(\bm{\Gamma}_{t}) \,
    g\!\left(c_{i0}^{2}\right)}_{\st}.
\end{equation}

Now we particularise the above general relation to the specific cases
we are interested in. Note that the dimensionless time $\uptau$ is
then used, instead
of $t$. First, we consider the following choice for  the function
$f$,
\begin{equation}
  f(\bm{\Gamma}_{\uptau})=\frac{1}{N}\sum_{i=1}^{N}\left(\frac{2}{d}
    c_{i}^{2}(\uptau)-1\right), \quad \mean{f(\bm{\Gamma}_{\uptau})}=
  \delta\theta(\uptau).
\end{equation}
Making use of Eq.~\eqref{eq:FDR-C-f-g-sum}, we have for the relaxation
function of the granular temperature
\begin{equation}\label{eq:FDR-temp}
  \phi_{\theta}(\uptau)=R_{2}(\uptau)\equiv\frac{1}{N}\sum_{i=1}^{N}\sum_{j=1}^{N}
  \mean{\left(\frac{2}{d}
    c_{i}^{2}(\uptau)-1\right)g(c_{j}^{2}(0))}_{\!\!\st}.
\end{equation}
Second, we consider the relaxation of the fourth moment. Therefore,
now we choose $f$ to be
\begin{align}
  f(\bm{\Gamma}_{\uptau})=\frac{1}{N}\sum_{i=1}^{N}\left(c_{i}^{4}(\uptau)-
  \mean{c^{4}}_{\st}\right), \quad
  \mean{f(\bm{\Gamma}_{\uptau})}=\delta\mean{c^{4}(\uptau)}, 
\end{align}
where $\mean{c^{4}}_{\st}$ is given by Eq.~\eqref{eq:c4-av-st}. Again,
we employ \eqref{eq:FDR-C-f-g-sum} to write
\begin{equation}\label{eq:FDR-c4}
  \phi_{\mean{c^{4}}}(\uptau)=
  R_{4}(\uptau)\equiv\frac{1}{N}\sum_{i=1}^{N}\sum_{j=1}^{N}
  \mean{\left(c_{i}^{4}(\uptau)-
  \mean{c^{4}}_{\st}\right)g(c_{j}^{2}(0))}_{\st}
\end{equation}

\section{Numerical simulations}\label{sec:numerical-sim-linear-response}

Now we compare our analytical predictions for the relaxation
function with numerical data obtained from the DSMC numerical
integration of the Boltzmann-Fokker-Planck equation. We have employed
a system of $N = 10^6$ hard discs ($d = 2$) of unit mass $m = 1$ and
unit diameter $\sigma = 1$, with the binary collision rule between
particles $i$ and $j$
\begin{equation}
  \bm{v}_{i}^{\prime}= \bm{v}_{i}- \frac{1+\alpha}{2}
  (\bm{\hat{\sigma}}\cdot \bm{v}_{ij}) 
  \bm{\hat{\sigma}}, \quad \bm{v}_{j}^{\prime}= \bm{v}_{j}+ \frac{1+\alpha}{2} (\bm{\hat{\sigma}}\cdot\bm{v}_{ij})
  \bm{\hat{\sigma}}.
\end{equation}
Above, $(\bm{v}_i,\bm{v}_j)$ are the precollisional velocities,
$(\bm{v}_i^{\prime},\bm{v}_j^{\prime})$ are the post-collisional
ones, $\bm{v}_{ij}\equiv \bm{v}_{i}-\bm{v}_{j}$ is the relative
velocity and $\bm{\hat{\sigma}}$ is the unit vector pointing from the centre
of particle $j$ to the centre of particle $i$ at the collision.
Moreover, in order to simulate the stochastic thermostat, the hard
discs are submitted to random kicks every $N_c = 500$ collisions. In
the kick, each component of the velocity of every particle is
incremented by a random number extracted from a Gaussian distribution
of variance $ \xi^2 \Delta t$, where $ \Delta t$ is the time interval
corresponding to the number of collisions $N_c$.

\subsection{Relaxation function of the temperature}

\begin{figure}
  \centering
  \includegraphics[width=1.625in]{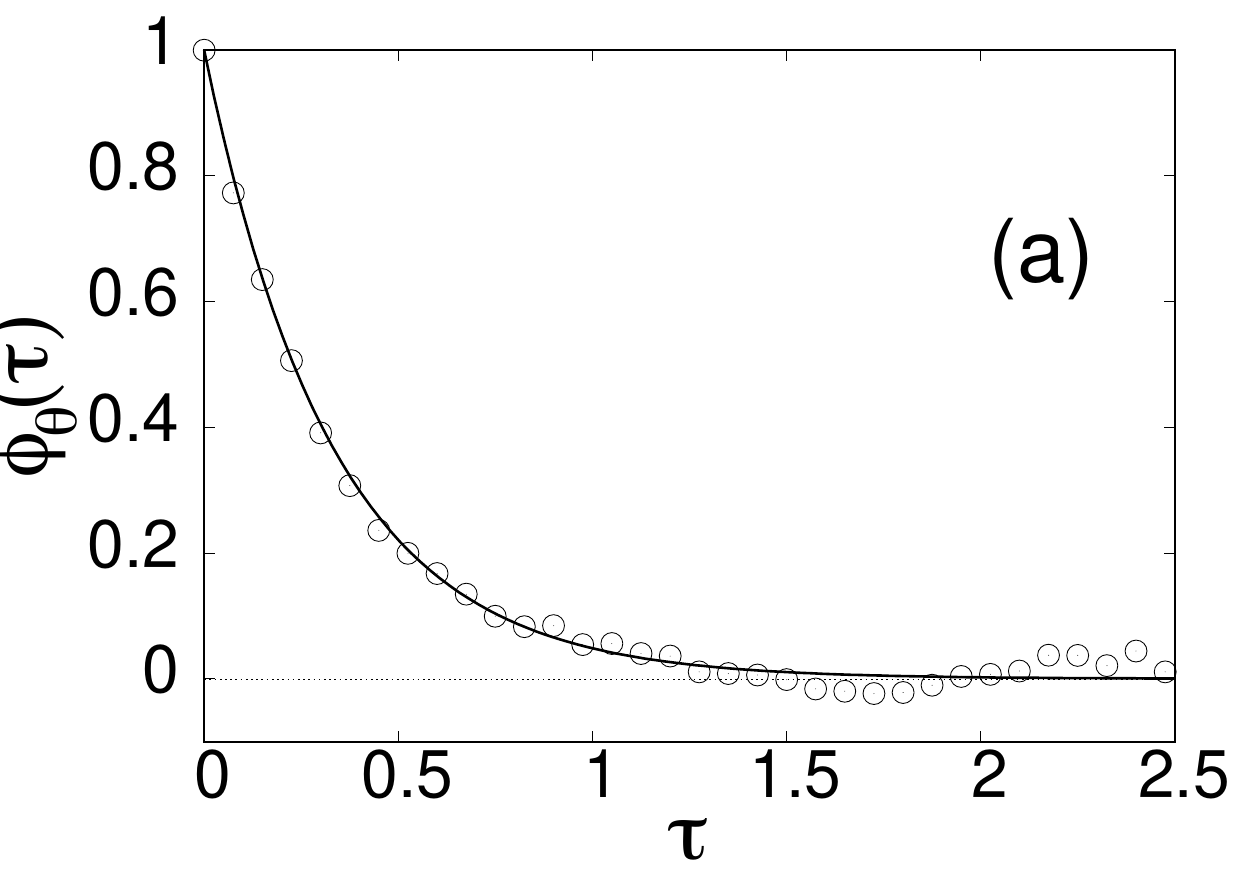}
  \includegraphics[width=1.625in]{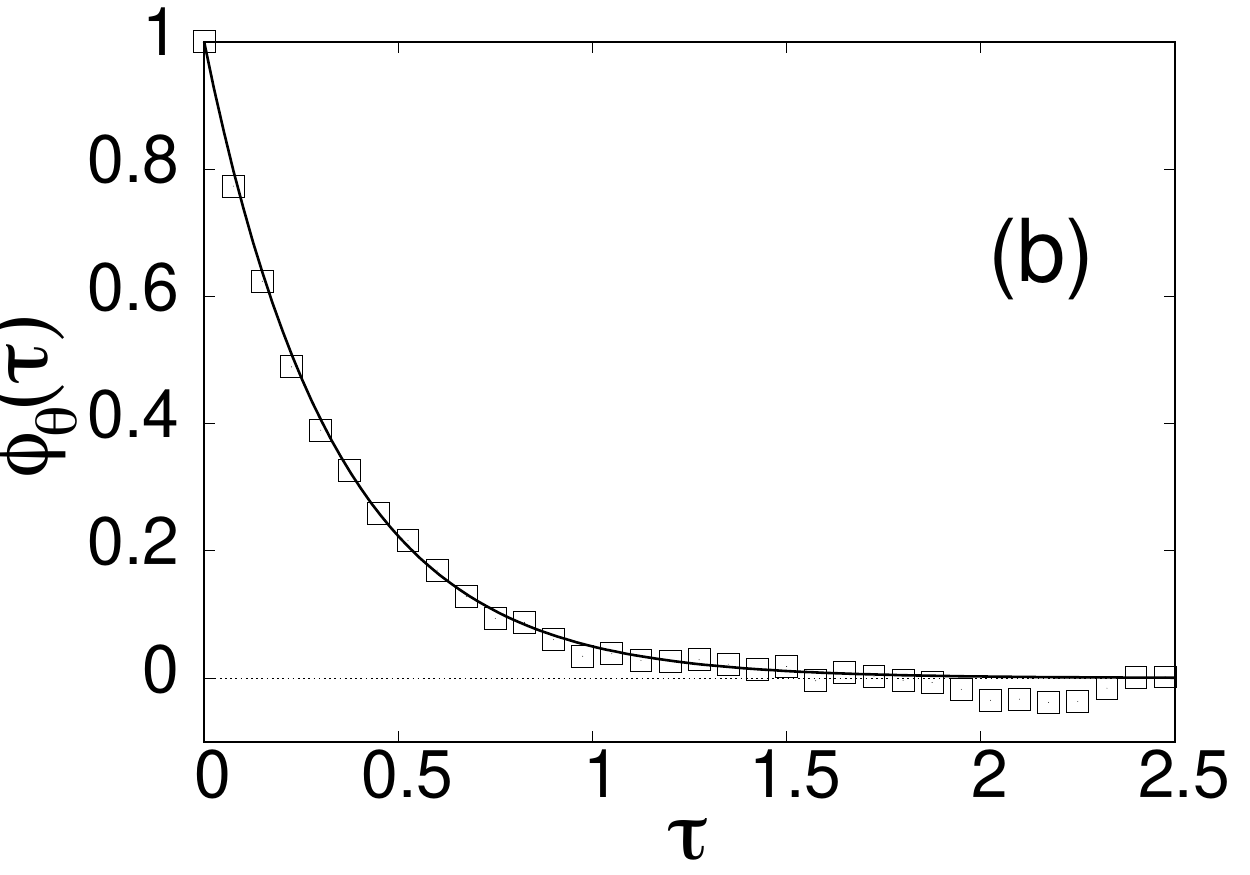}
  \includegraphics[width=1.625in]{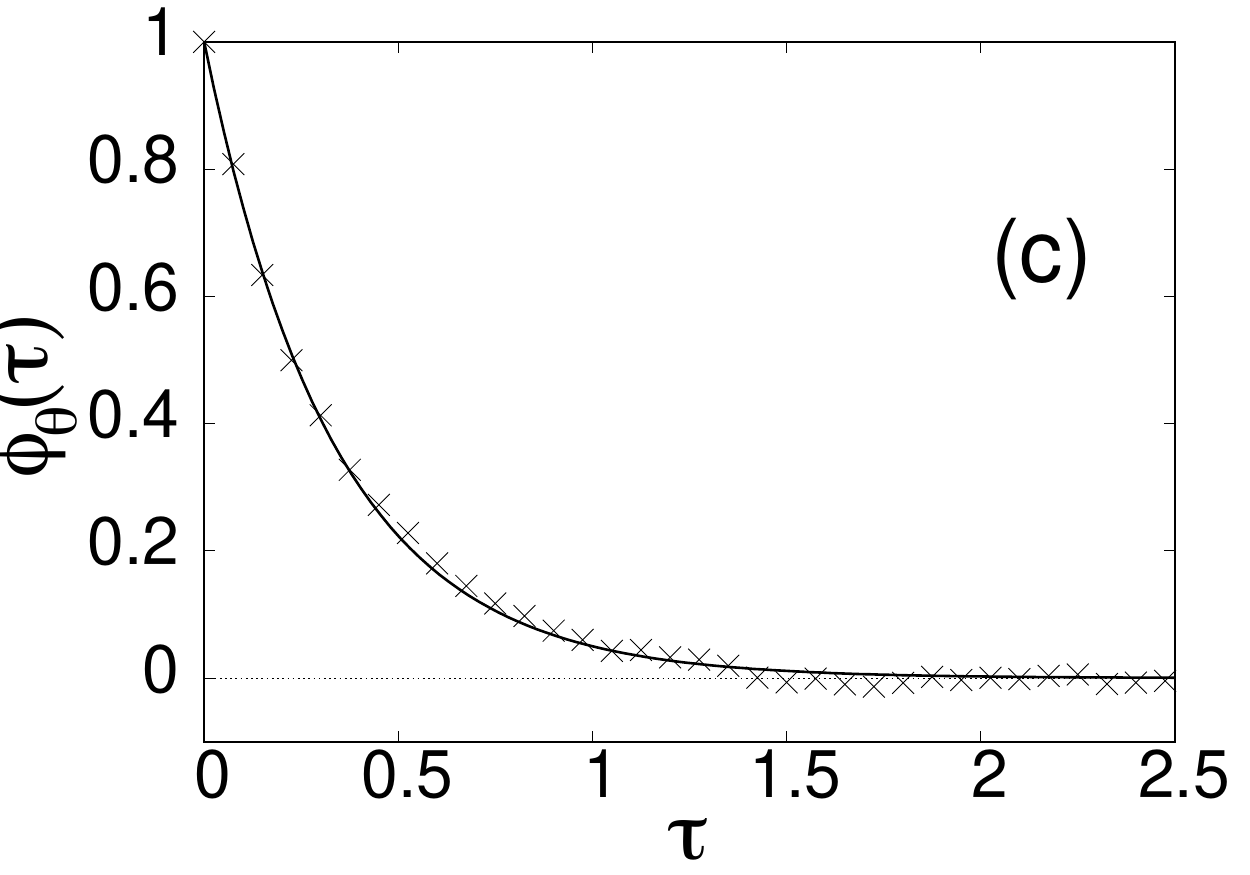}
  \includegraphics[width=1.625in]{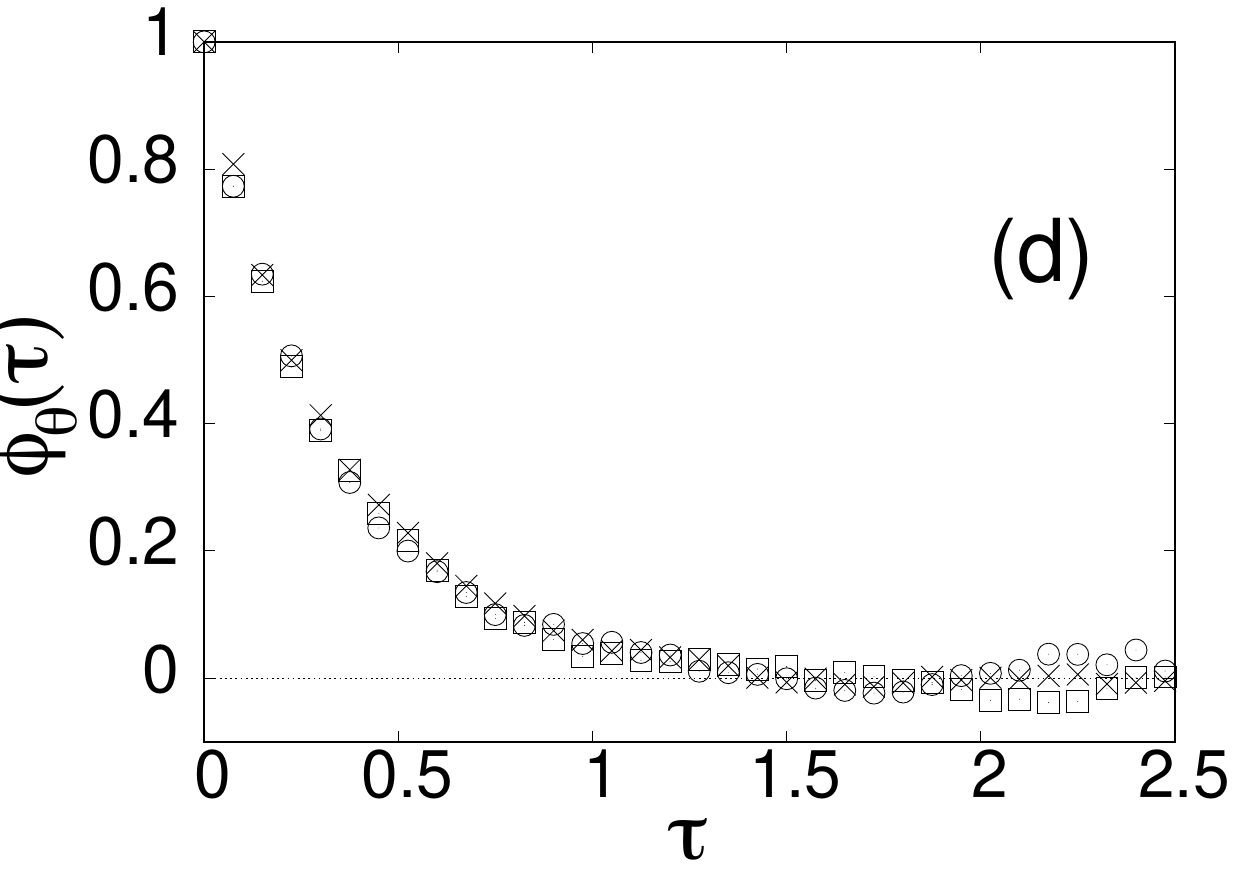}
  \caption{Relaxation function of the granular temperature as a
    function of the scaled time $\uptau$. Panels (a), (b), and (c)
    display $\phi_{\theta}(\uptau)$ for three different values of the
    restitution coefficient, namely $\alpha=0.3$, $\alpha=0.7$ and
    $\alpha=0.9$, respectively---symbols correspond to simulation
    results in a system of $10^{6}$ particles averaged over $10$ runs, and lines to the analytical
    prediction~\eqref{eq:phi11}. The three simulation curves are
    plotted together in panel (d), where it is clearly seen that they
    superimpose. Therefore, the linear relaxation of the granular
    temperature to the corresponding NESS is ``universal'' in the
    $\uptau$ scale.}
  \label{fig:T-relax-functions}
\end{figure}

In particular, we show the relaxation functions corresponding to
$\alpha=0.3$ ($>\alpha_{c}$), $\alpha=0.7$ ($\simeq\alpha_{c}$), and
$\alpha=0.9$ ($<\alpha_{c}$)---see panels (a) to (c) of
Fig.~\ref{fig:T-relax-functions}.  The symbols represent simulation
results from a NESS at $(\xi+\delta \xi)^2=0.205$ to the NESS
corresponding at $\xi^2=0.2$, while the solid lines represent our
theoretical prediction \eqref{eq:phi11}.  The agreement is excellent,
which comes as no surprise because the first Sonine approximation is
known to give a good description of the granular gas
dynamics---although the majority of the studies have been done in the
non-linear
regime~\cite{garcia_de_soria_universal_2012,prados_kovacs-like_2014,trizac_memory_2014}.

\begin{figure}
  \centering
  \includegraphics[width=2.75in]{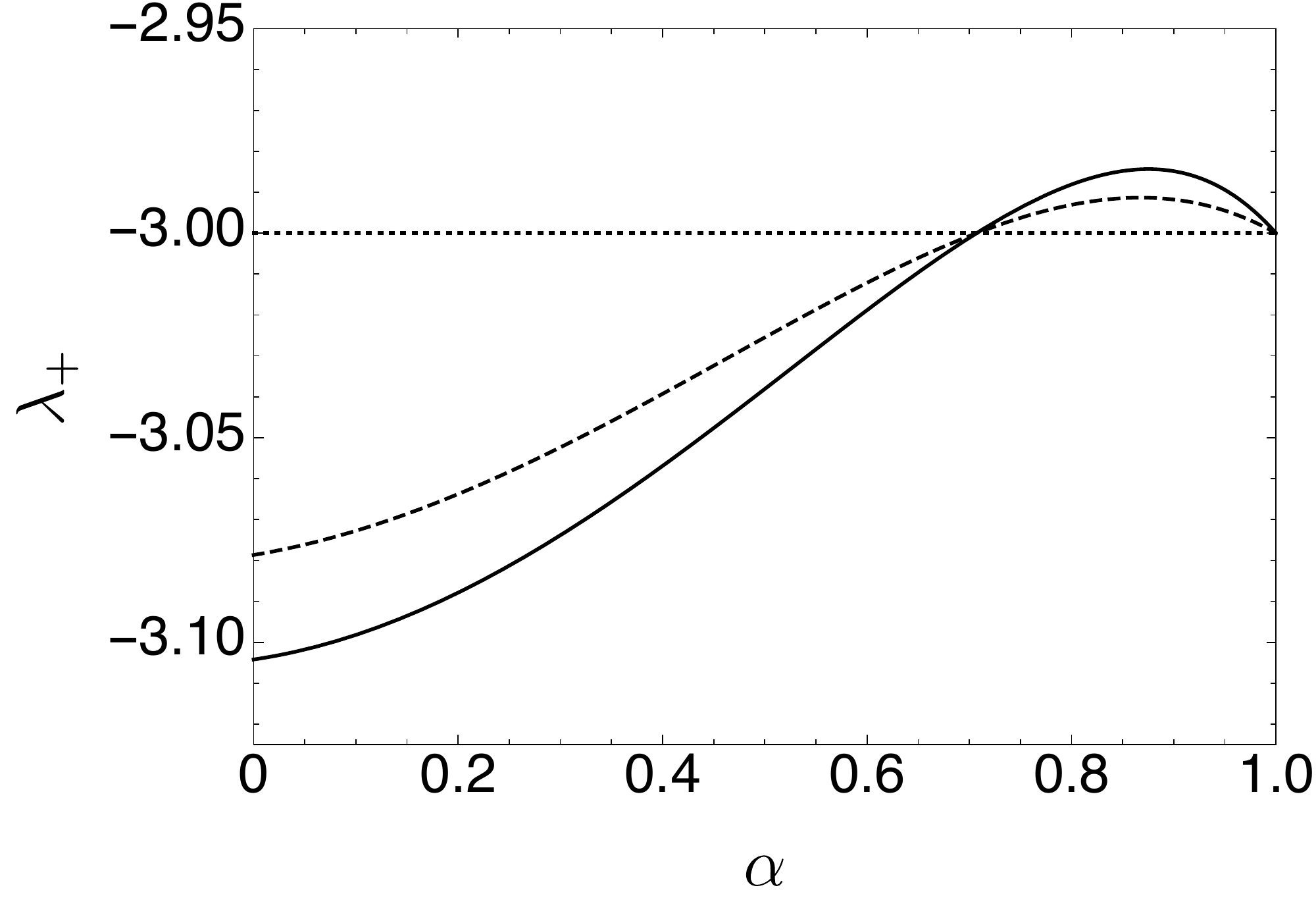}\\
  \includegraphics[width=2.75in]{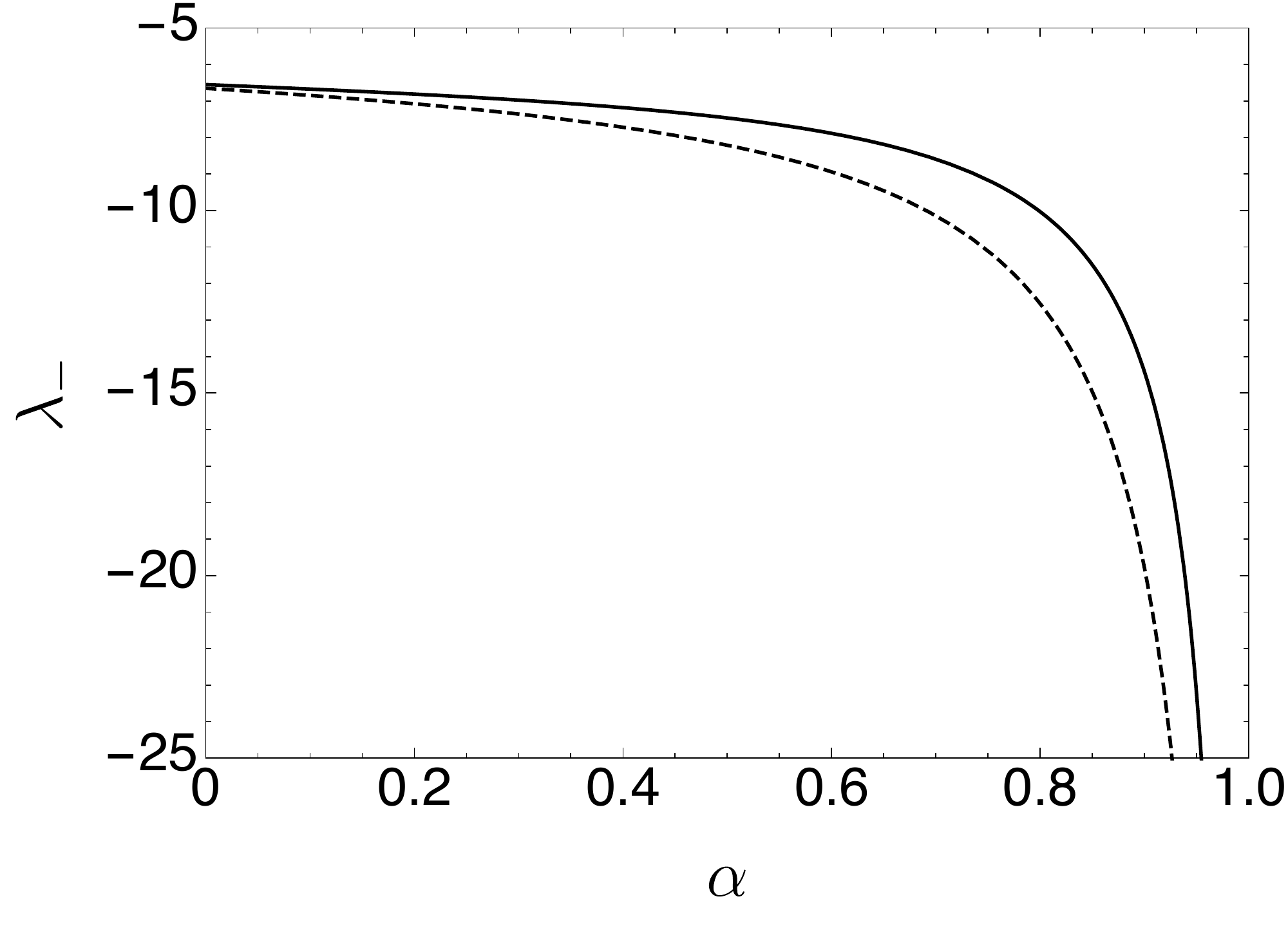}
  \caption{\label{fig:lambda+-and-lambda-}Relaxation rates
    $\lambda_{\pm}$ as a function of the restitution
    coefficient. Again, both $d=3$ (solid line) and $d=2$ (dashed) are
    plotted. The largest rate $\lambda_{+}$ is always close to $-3$
    throughout. The smaller rate $\lambda_{-}$ has a stronger
    dependence on $\alpha$, diverging in the elastic limit $\alpha\to
    1^{-}$.
  }
\end{figure}

In principle, Eq.~\eqref{eq:phi11} tells us that the relaxation of the
granular temperature to the NESS is non-exponential. However, the
smallness of the coefficient $a_{-}$ throughout the whole $\alpha$
range entails that, from a practical point of view, the relaxation is
very well fitted by a single exponential for all values of the
restitution coefficient. Moreover, the relaxation functions for
different values of $\alpha$ superimpose, when plotted as a function
of the scaled time $\uptau$~\footnote{They do not superimpose as a
  function of the real time $t$. The definition of the scaled time in
  Eq.~\eqref{eq:scaled-variables-def} involves $\alpha$ because the
  parameter $\zeta_{0}$ is proportional to $1-\alpha^{2}$. Therefore,
  the smaller the inelasticity $1-\alpha$ is, the slower the
  relaxation in real time becomes.}. This is neatly observed in panel
(d) of Fig.~\ref{fig:T-relax-functions}, where we have put together
the simulation results from the three previous panels. This stems from
the weak $\alpha$-dependence of the eigenvalue $\lambda_{+}$, which is
close to $-3$ for all values of $\alpha$, as predicted by
Eq.~\eqref{eq:lambda+-expansion} and shown in panel (a) of
Fig.~\ref{fig:lambda+-and-lambda-}. Therein, it is observed that the
maximum deviation from $-3$ is of the order of $3\%$, for the largest
value of $a_{2}^{\st}$---i.e. in the completely inelastic limit
$\alpha=0$. For completeness, we show $\lambda_{-}$ in panel (b) of
the same figure.

\begin{figure*}
  \centering
  \includegraphics[width=3.25in]{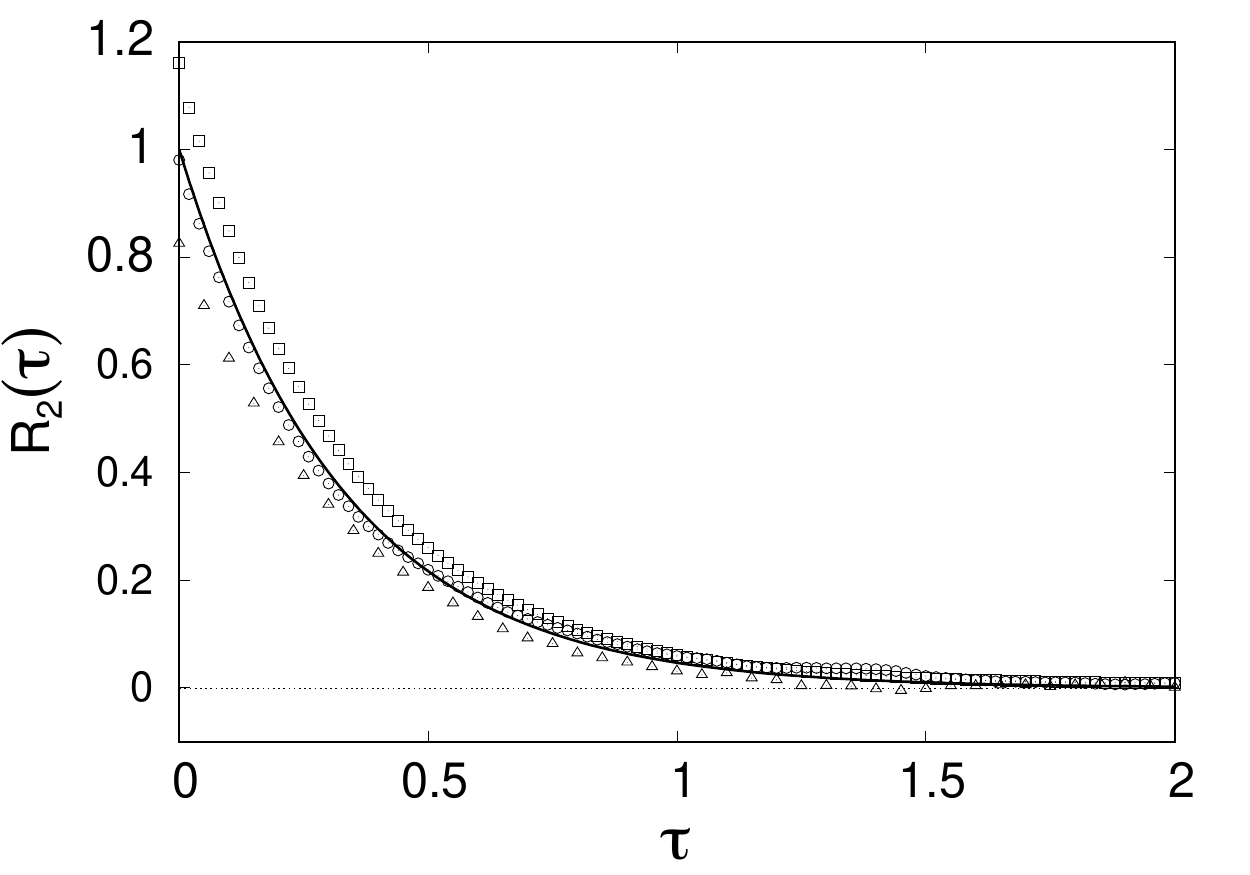}
  \includegraphics[width=3.25in]{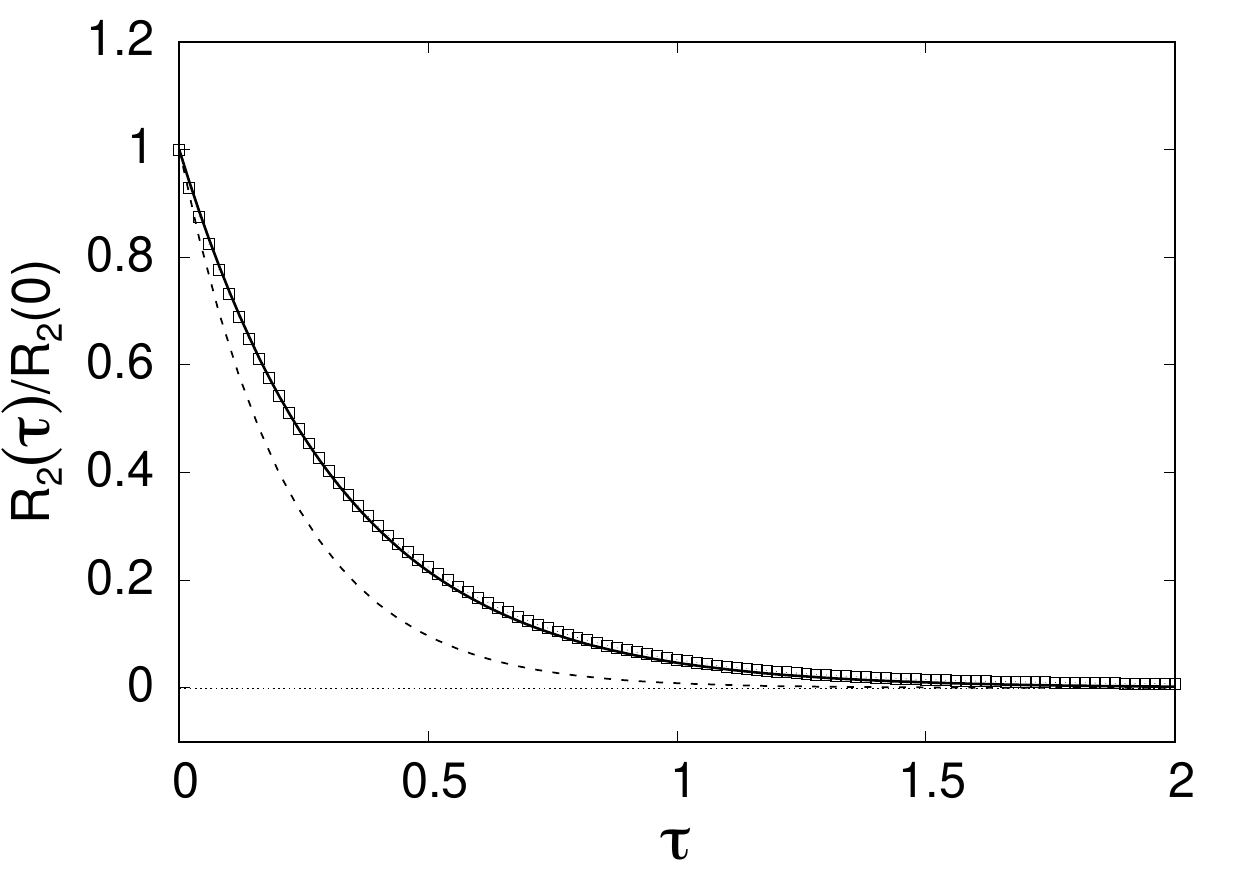}
  \caption{Check of the FDR for the relaxation function of the
    granular temperature. The left panel shows the correlation
    function $R_{2}(\uptau)$ for three different values of the
    restitution coefficient $\alpha$: from top to bottom, $\alpha=0.3$
    (squares), $0.7$ (circles), and $0.9$ (triangles). Also shown is
    the theoretical prediction for the direct relaxation function of
    the granular temperature $\phi_{\theta}(\uptau)$ (line). As
    discussed in the text, the decay of $\phi_{\theta}$ is well
    described by $R_{2}$, although the initial values do not exactly
    match. This is illustrated in the right panel, where we plot
    $R_{2}(\uptau)/R_{2}(0)$ for the specific case $\alpha$=0.3.
    Therein, $R_{2}(\uptau)/R_{2}(0)$ is compared with both
    $\phi_{\theta}(\uptau)$ (solid line) and the diagonal part of the
    correlation $R_{2}^{\diag}(\uptau)$ (dashed line). It is clearly
    observed that the diagonal part equals unity for $\uptau=0$, but
    it decays faster than the whole correlation $R_{2}$: roughly, its
    relaxation time is divided by two.}
  \label{fig:T-FDR-corr-function}
\end{figure*}

Now we proceed to evaluate the relaxation function of the temperature
by using the FDR relation \eqref{eq:FDR-temp}. The correlation
function on the rhs has been numerically evaluated from the DSMC data,
employing system with $N=10^{4}$--$10^{6}$ hard discs and averaging
over a long trajectory of duration $\Delta\uptau=400$--$6000$,
depending on the value of $\alpha$.  We compare the numerics with
Eq.~\eqref{eq:FDR-temp} for $\phi_{\theta}$ in
Fig.~\ref{fig:T-FDR-corr-function}. On the one hand, we observe in the
left panel that there appears a discrepancy at the initial time. The
value stemming from the FDR relation deviates from unity in general:
it is larger (smaller) for $\alpha=0.3$ $\alpha=0.9$, whereas it
equals one for $\alpha=0.7\simeq \alpha_{c}$. On the other hand, the
right panel shows that this
discrepancy disappears if we normalise the correlation function
$R_{2}(\uptau)$ with its initial value, i.e. if we compare
$\phi_{\theta}(\uptau)$ with $R_{2}(\uptau)/R_{2}(0)$. Therefore, the
correlation $R_{2}(\uptau)$ correctly predicts the decay,
$R_{2}(\uptau)=R_{2}(0)\phi_{\theta}(\uptau)$, but not the initial
value.

In light of the above, it is worth asking the reason behind the
discrepancy for the initial time. The correlation function includes
both \textit{diagonal} terms ($i=j$) and \textit{non-diagonal} terms
($i\ne j$) in its double sum, thus we split it accordingly into
\begin{subequations}
\begin{align}
  R_{2}(0)&=R_{2}^{\diag}(0)+R_{2}^{\ndiag}(0), \\
  R_{2}^{\diag}(0)&=\frac{1}{N}\sum_{i=1}^{N}
  \mean{\left(\frac{2}{d}
                         c_{i}^{2}-1\right)g(c_{i}^{2})}_{\!\!\st} \nonumber \\
  & =\mean{\left(\frac{2}{d}
    c_{1}^{2}-1\right)g(c_{1}^{2})}_{\!\!\st}, \\
  R_{2}^{\ndiag}(0)&=\frac{1}{N}\sum_{i=1}^{N}\sum_{j\ne i}
  \mean{\left(\frac{2}{d}
  c_{i}^{2}-1\right)g(c_{j}^{2})}_{\!\!\st} \nonumber \\
  & =(N-1)\mean{\left(\frac{2}{d}
    c_{1}^{2}-1\right)g(c_{2}^{2})}_{\!\!\st}.
\end{align}
\end{subequations}
Note that the above one-time averages are done in the NESS and are
thus time-independent, so we have written $c(0)\to c$. Employing the
factorisation assumption \eqref{eq:propag-chaos} and
the first Sonine approximation \eqref{eq:P1s-Sonine} for
$P_{1}^{\st}$, we obtain that
\begin{subequations}
  \begin{align}
  R_{2}^{\diag}(0)&=\mean{\left(\frac{2}{d}
    c^{2}-1\right)g(c^{2})}_{\!\!\st}=1, \\
  R_{2}^{\ndiag}(0)&=(N-1) \mean{\left(\frac{2}{d}
    c^{2}-1\right)}_{\!\st}\mean{g(c^{2})}_{\st}=0.
  \end{align}
\end{subequations}

We have also evaluated $R_{2}^{\diag}(\uptau)$ and
$R_{2}^{\ndiag}(\uptau)$ from the DSMC data. We have always got
$R_{2}^{\diag}(0)\simeq 1$, whereas in general
$R_{2}^{\ndiag}(0)\ne 0$: it is the non-diagonal terms that the
discrepancy at the initial time stems from. The physical reason behind
this is the $O(N^{-1})$ correlations that are completely neglected
when writing Eq.~\eqref{eq:propag-chaos}. In order to account for this
behaviour, it is necessary to consider the two-particle time
correlations, which is outside the scope of this work~\footnote{Only
  the ``static'', time-independent, correlations in the NESS of the
  uniformly heated granular gas have been previously
  investigated~\cite{garcia_de_soria_energy_2009}.}. As shown in the
right panel of Fig.~\ref{fig:T-FDR-corr-function},
$R_{2}^{\diag}(\uptau)$ decays faster than the whole correlation
function $R_{2}(\uptau)$, the relaxation time of the latter
approximately doubles that of the former.

It must be remarked that the discrepancy between the relaxation
function and the correlation is not a consequence of the first Sonine
approximation. One can easily incorporate the contribution from higher
cumulants into $R_{2}(\uptau)$, for instance the following term that
includes the sixth-cumulant $a_{3}$ and the third Sonine polynomial
$S_{3}(x)$. The only change is that now $g(x)$ reads
\begin{equation}
  g(x)=-\frac{d}{2}+x-x
  \frac{a_{2}^{\st}S_{2}'(x)+a_{3}^{\st}S_{3}'(x)}
  {1+a_{2}^{\st}S_{2}(x)+a_{3}S_{3}(x)}.
\end{equation}
We have checked that the DSMC results for the correlation function remain
basically unaltered.

\subsection{Relaxation function of the fourth-moment}

\begin{figure}
  \centering
  \includegraphics[width=3.25in]{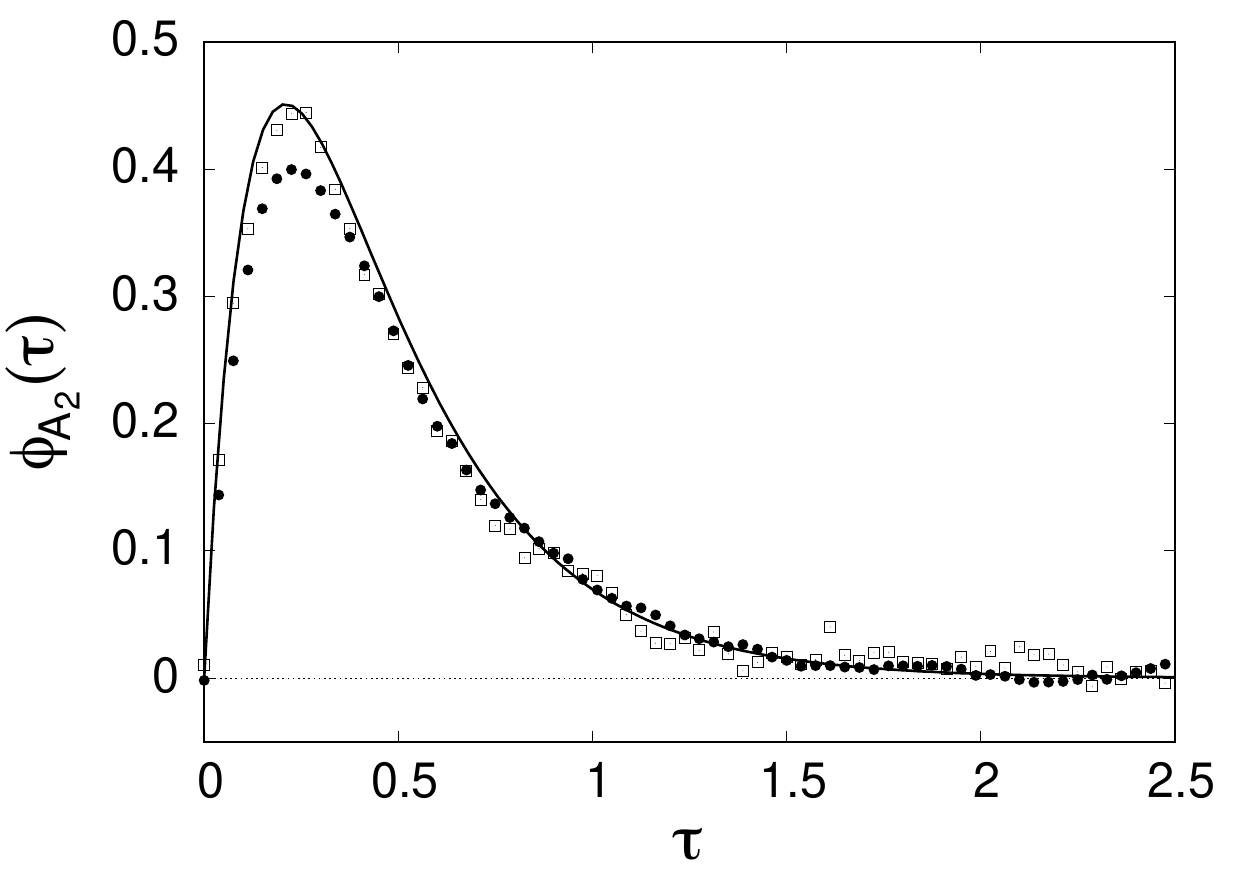}
  \caption{\label{fig:relax-A2} Relaxation function of the scaled
    excess kurtosis $A_{2}=a_{2}/a_{2}^{\st}$ as a function of
    $\uptau$, for $\alpha=0.3$. In the simulations, the driving is
    instantaneously changed from $(\xi+\delta \xi)^2=0.25$ (squares)
    and $(\xi+\delta \xi)^2=0.35$ (solid circles) to $\xi^{2}=0.2$ at
    $\uptau=0$. Despite the jumps being quite large, the agreement
    with the theoretical prediction, as given by Eq.~\eqref{eq:phi1-2}
    (line), is remarkably good.}
\end{figure}
Simulation curves for the time evolution of $A_{2}$ are compared with
the theoretical prediction \eqref{eq:phi1-2} in
Fig.~\ref{fig:relax-A2}. All curves correspond to a system with
restitution coefficient $\alpha=0.3$ and have been averaged over $100$
runs. Two experiments in which the driving intensity is decreased from
$\xi+\delta\xi$ to $\xi$ are considered. In both cases, we have that
$\xi^{2}=0.2$, but with two different values of the jump,
$(\xi+\delta \xi)^2=0.25$ (squares) and $(\xi+\delta \xi)^2=0.35$
(solid circles).

\begin{figure}
  \centering
  \includegraphics[width=3.25in]{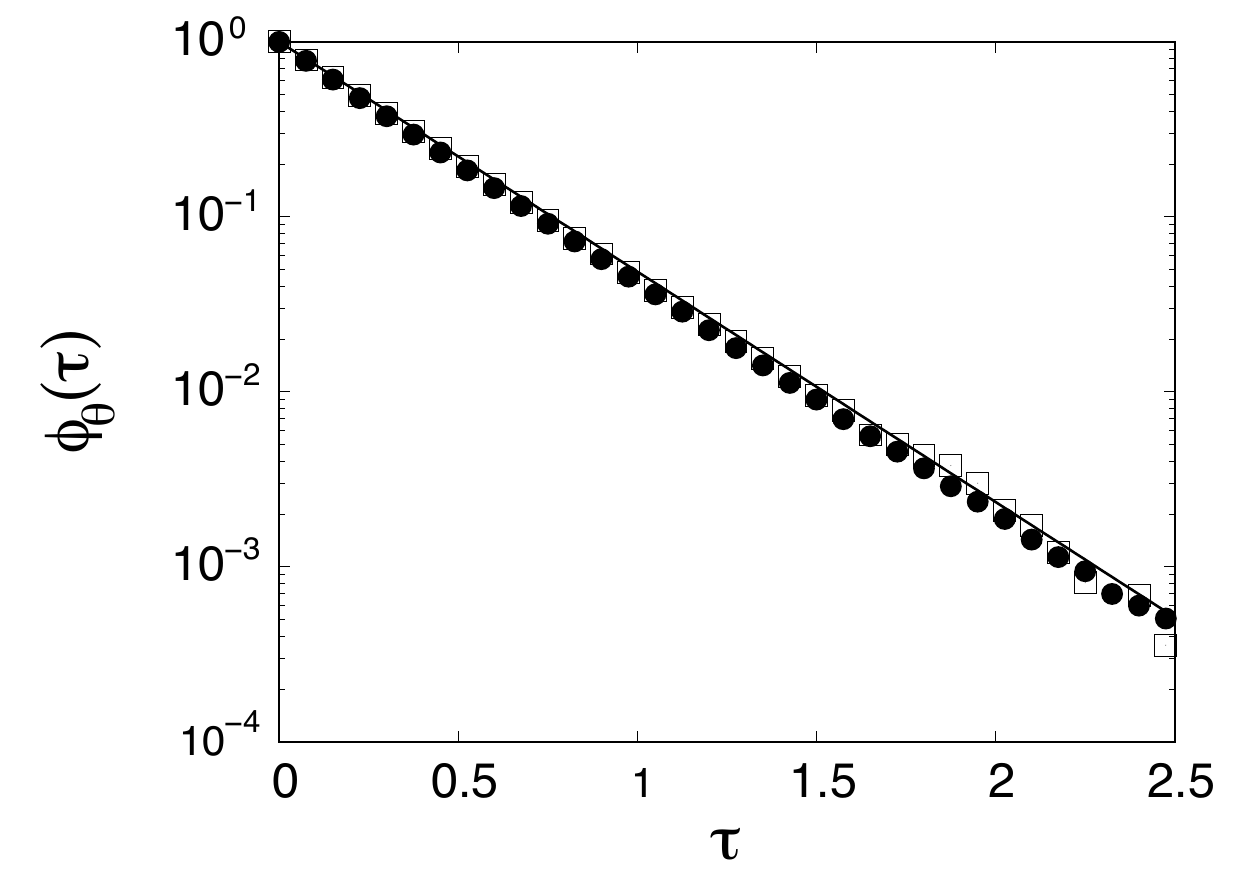}
  \caption{\label{fig:relax-T-larger-jumps}Relaxation function of the
    granular temperature for the same situation considered in
    Fig.~\ref{fig:relax-A2}. Note the logarithmic scale on the
    vertical axis. Although the jumps in the driving are quite large
    ---relative size of $10$--$30\%$,
    the linear response prediction (solid line) fits perfectly well
    the simulation results (symbols). The coding for the different
    lines is the same as in Fig.~\ref{fig:relax-A2}.}
\end{figure}

Fluctuations in $A_{2}$ are larger than those of the
granular temperature and thus it is necessary to consider larger jumps
in the driving, roughly one order of magnitude larger than those
employed in Fig.~\ref{fig:T-relax-functions}. Still, linear response
theory works pretty well: for $(\xi+\delta \xi)^2=0.25$,
$\delta\xi/\xi\simeq 0.12$ but the agreement is almost perfect; for
$(\xi+\delta \xi)^2=0.35$, $\Delta\xi/\xi\simeq 0.3$ and the theory
only slightly overestimates the relaxation function
$\phi_{A_2}^{\uptau}$. For these larger jumps in the driving, we have also
looked into the relaxation of the granular temperature. This is done
in Fig.~\ref{fig:relax-T-larger-jumps}, the agreement between the
simulation curves and the linear response result is even better than
that of the excess kurtosis.

\begin{figure}
  \centering
  \includegraphics[width=3.25in]{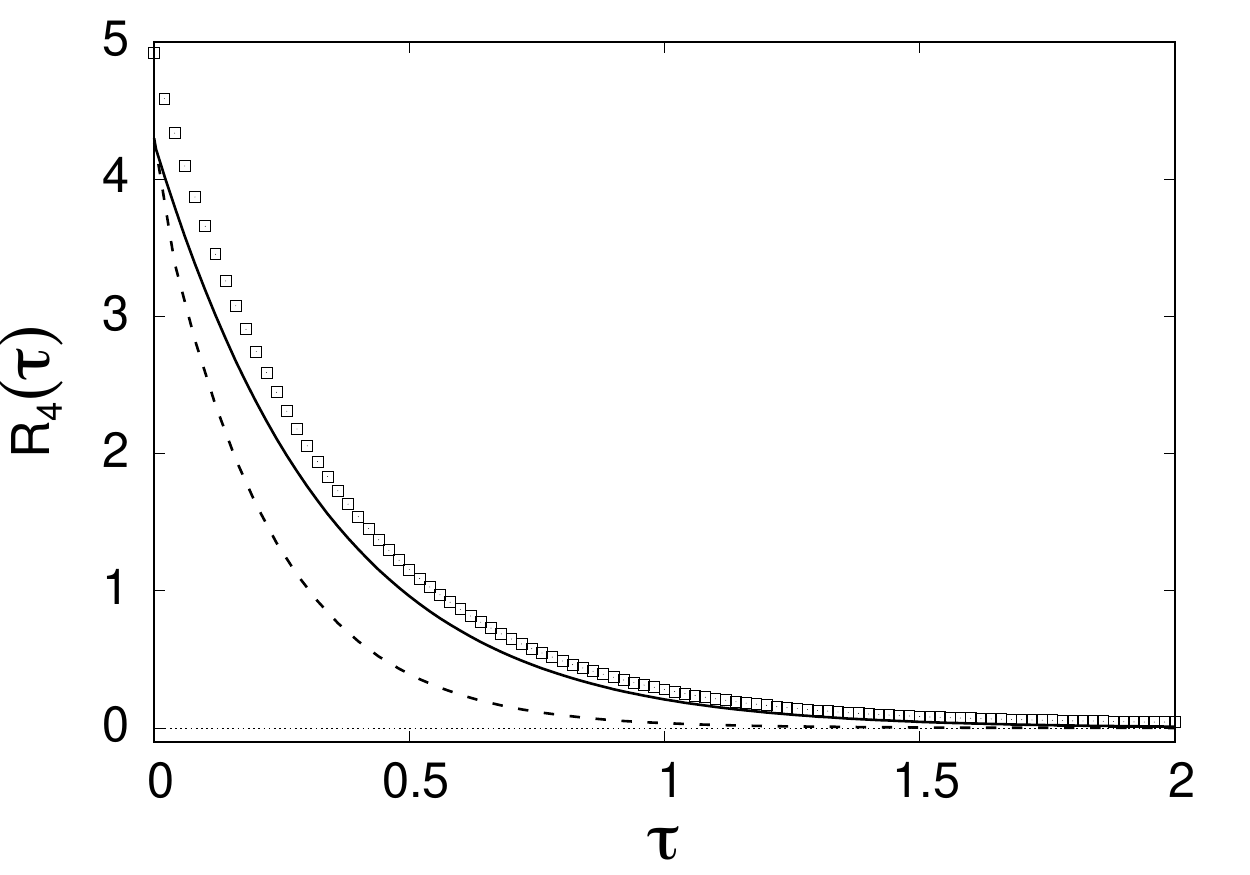}
  \caption{Check of the FDR relation for the relaxation function of
    the fourth-moment. The squares represent the correlation function
    $R_4(\tau)$ for $\alpha=0.3$, measured in DSMC simulations, while
    the solid line is the theoretical prediction given by
    Eq.~\eqref{eq:phic4-direct}. The decay of $\phi_{\mean{c^{4}}}$ is
    well described by $R_4$, although their initial values do not
    exactly match. On the other hand, the diagonal part of the
    correlation $R_{4}^{\diag}(\uptau)$ (dashed line) correctly gives
    the initial value $\phi_{\mean{c^{4}}} (0)$, but it decays faster
    than the whole correlation $R_4$.}
  \label{fig:c4-FDR-corr-function}
\end{figure}

We have also checked the accuracy of the FDR relation
\eqref{eq:FDR-c4} for the relaxation function $\phi_{\mean{c^{4}}}$
of the fourth moment. We have numerically computed the time
correlation function $R_{4}(\uptau)$ from the DSMC data, the values
for the simulation parameters are the same as for the temperature. In
Fig.~\ref{fig:c4-FDR-corr-function}, for $\alpha=0.3$, the numerics
(squares) is compared with the theoretical prediction for
$\phi_{\mean{c^{4}}}$, as given by Eq.~\eqref{eq:phic4-direct} (solid
line). Similarly to the case of the temperature relaxation, the
correlation function $R_4(\uptau)$ correctly predicts the decay but
not the initial value of $\phi_{\mean{c^{4}}}$. Again, this
discrepancy comes out from the non-diagonal terms of the
correlation function, as the evaluation of the diagonal part
$R_{4}^{\diag}(\uptau)$ (dashed line) shows.

\section{The Kovacs-like experiment with two jumps in the driving}\label{sec:kovacs-exp}

In this section, we analyse a more complex experiment, which was first
done by Kovacs to investigate the glassy response of
polymers~\cite{kovacs_transition_1963,kovacs_isobaric_1979}. The main
idea is the following: first, the system under study is ``aged'' by
following a certain protocol. After this aging stage, the system has
 values of its macroscopic variables equal to those in the
steady state. However, the macroscopic variables do not remain flat
but display a ``hump''. This is the Kovacs effect, which tells us that
there are other, non-macroscopic, variables that also have to be taken
into account to understand the relaxation properties of the system.

Here, we describe the Kovacs-like experiment in terms of the variables
of our model~\footnote{Aside from the original papers by
  Kovacs~\cite{kovacs_transition_1963,kovacs_isobaric_1979}, the
  account of the original experiment can be found in many places, see
  for example Refs.~\cite{scherer_relaxation_1986,
    scherer_theories_1990,prados_kovacs_2010}.}.  Instead of letting
the system relax directly from the NESS for $\xi+\delta\xi$ to
that for $\xi$, at $t=0$ we suddenly change the driving intensity from
$\xi+\delta\xi$ to a lower driving $\xi-\delta\xi'<\xi$. Then,
the system starts to relax to the NESS for $\xi-\delta\xi'$, at
which $T_{\st}(\xi-\delta\xi')=T_{\st}(\xi)-\delta
T'<T_{\st}(\xi)$. At some time---which we call henceforth the
waiting time $\uptau_{w}$---the instantaneous value of the granular
temperature $T(\uptau)$  equals $T_{\st}(\xi)$. At
$\uptau=\uptau_{w}$, we suddenly change the driving intensity from
$\xi-\delta\xi'$ to $\xi$. This two-jump procedure in the
driving characterises the Kovacs protocol.

\begin{figure}
  \centering
    \includegraphics[width=3.25in]{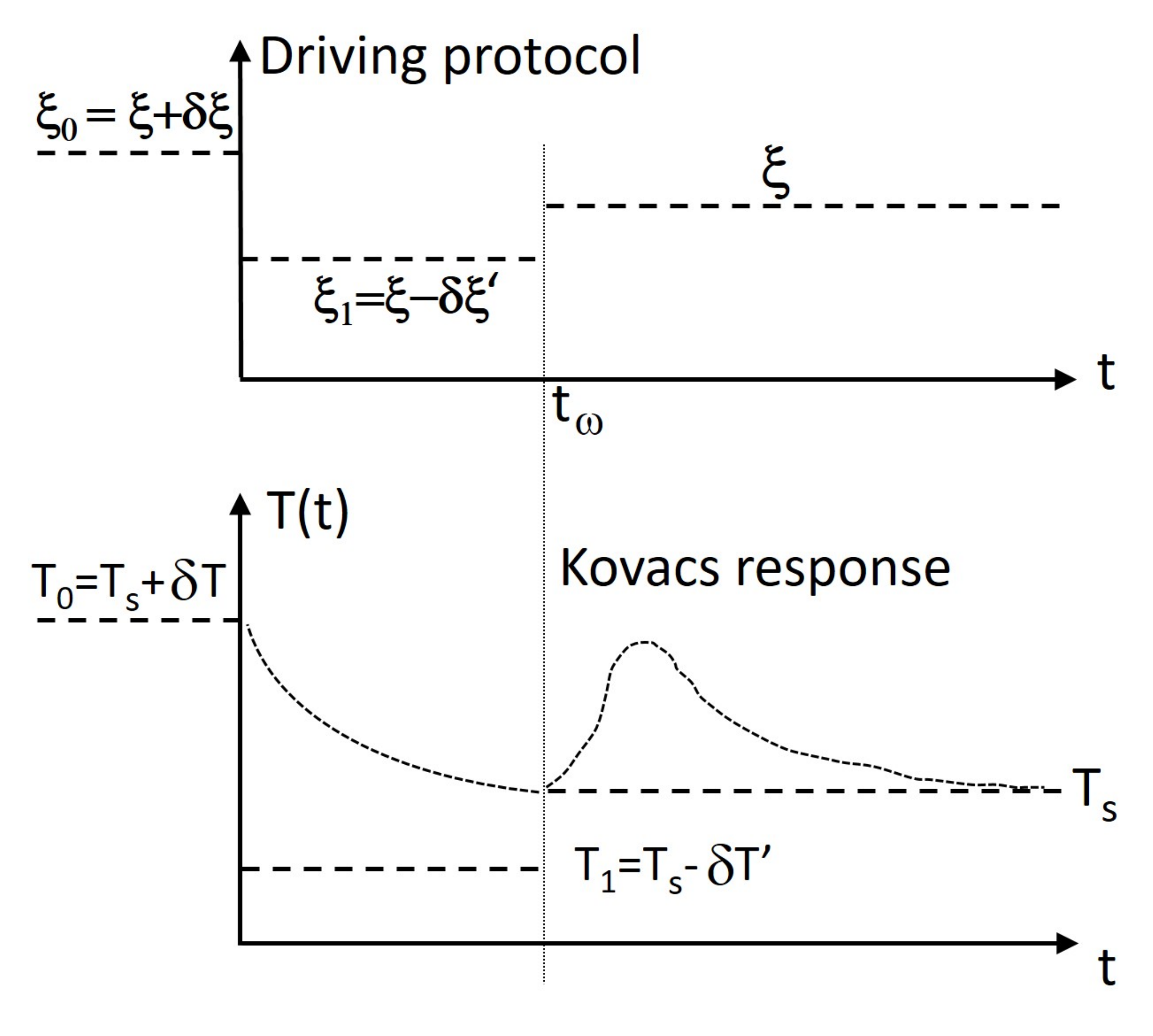}
    \caption{Qualitative plot of the Kovacs protocol.  The system
      starts from a NESS with granular temperature
      $T_{0}=T_{\st}+\delta T$, corresponding to a driving
      $\xi_{0}=\xi+\delta\xi$. In the waiting time window, the driving
      is suddenly decreased to $\xi_{1}=\xi-\delta\xi'<\xi$ and the
      granular temperature approaches the corresponding steady state
      value $T_{1}=T_{\st}-\delta T'<T_{\st}$. At the waiting time,
      the granular temperature equals its steady value for the driving
      $\xi$. However, for longer times the granular temperature does
      not remain flat: it displays a non-monotonic behaviour, the
      Kovacs hump. The normal case is depicted, in which the
      temperature displays a maximum, but anomalous behaviour---a
      minimum instead of a maximum---can also be observed. Note that
      $\delta T$ and $\delta T'$ are defined in such a way that they
      are both positive.}
  \label{fig:protocol}
\end{figure}

A qualitative plot of the Kovacs protocol is shown in
Fig.~\ref{fig:protocol}. Specifically, we have considered the usual
case in which $\delta\xi$ and $\delta\xi'$ are both positive and
therefore $T_{\st}+\delta T>T_{\st}>T_{\st}-\delta T'$---sometimes
termed the \textit{cooling} protocol. The Kovacs effect is brought to
the fore if $T(\uptau)$ does not remain flat for $\uptau>\uptau_{w}$,
signalling that, actually, the system is not in the NESS for the
driving intensity $\xi$ at $\uptau=\uptau_{w}$. If
$T(\uptau)-T_{\st}>0$, the Kovacs effect is normal---this is the
situation that is always found in molecular
systems~\cite{kovacs_isobaric_1979,kovacs_transition_1963,prados_kovacs_2010},
and the one plotted in the figure. If $T(\uptau)-T_{\st}<0$, the
Kovacs effect is anomalous, a possibility that has recently been
reported in some athermal
systems~\cite{prados_kovacs-like_2014,trizac_memory_2014,kursten_giant_2017}.

In the following, we analyse the Kovacs hump in linear response.  Note
that the theoretical approach presented here essentially differs from
that in Refs.~\cite{prados_kovacs-like_2014,trizac_memory_2014}, where
the non-linear case---large jumps in the driving---was
investigated. Therein, a perturbative analysis in powers of
$a_{2,\st}$ made it possible to derive an analytical expression for
the Kovacs hump, but only when the system freely cooled in the waiting
time window~\footnote{For $\xi_{1}=0$, the system reached the
  homogeneous cooling state and $A_{2}$ the specific value
  $a_{2}^{\hcs}/a_{2}^{\st}$ in the waiting time window, after a quite
  short transient.}. The linear response approach developed below,
although limited to small jumps in the driving, improves our
understanding of the Kovacs effect by connecting the shape of the hump
with the properties of the direct relaxation function.

The evolution of the temperature for $\uptau\geq\uptau_{w}$ directly
follows from the general linear response scheme in
Ref.~\cite{plata_kovacs-like_2017}. Namely, it is given by
\begin{equation}\label{eq:deltaT-Kovacs}
  \delta
  T(\uptau)=\left(T_{0}-T_{1}\right)\left[\phi_{\theta}(\uptau)-
    \phi_{\theta}(\uptau_{w})\phi_{\theta}(\uptau-\uptau_{w})\right],
\end{equation}
where $\phi_{\theta}$ is the direct relaxation function (i.e., for the
one jump experiment), normalised in the sense that
$\phi_{\theta}(\uptau=0)=1$. Substitution of Eq.~\eqref{eq:phi11} into 
\eqref{eq:deltaT-Kovacs} results, after a little bit of algebra,  in
\begin{align}
  \delta
  \theta(\uptau)=\left(\delta\theta+\delta\theta'\right)
  &a_{+}a_{-}\left(e^{\lambda_{+}\uptau_{w}}-e^{\lambda_{-}\uptau_{w}}\right)\nonumber\\ 
  &\times\left(e^{\lambda_{+}(\uptau-\uptau_{w})}-e^{\lambda_{-}(\uptau-\uptau_{w})}\right),
    \label{eq:Kovacs-delta-theta-tau}
\end{align}
where we have employed the dimensionless temperature jumps
\begin{equation}\label{eq:deltatheta-deltatheta'}
  \delta\theta\equiv \frac{\delta T}{T_{\st}}=\frac{T_{0}-T_{\st}}{T_{\st}},
  \quad \delta\theta'\equiv \frac{\delta T'}{T_{\st}}=\frac{T_{\st}-T_{1}}{T_{\st}},
\end{equation}
such that $T_{0}-T_{1}=T_{\st}(\delta\theta+\delta\theta')$. The time
evolution of $\delta\theta(\uptau)$ is non-monotonic: it is the
amplitude $a_{-}$ of the second mode of the direct relaxation function
$\phi_{\theta}$ that controls the sign of the Kovacs response, since
the remainder of the terms on the rhs of
Eq.~\eqref{eq:Kovacs-delta-theta-tau} are positive definite.  Normal
behaviour, i.e.  $\delta\theta(\uptau)\geq 0$, is obtained for
$a_{-}>0$, i.e. for $\alpha>\alpha_{c}$---see Eq.~\eqref{eq:sgn-a2}
and Fig.~\ref{fig:a+-and-a-}. Anomalous behaviour,
i.e. $\delta\theta(\uptau)\leq 0$, is obtained for $a_{-}<0$, i.e. for
$\alpha<\alpha_{c}$.

The above discussion entails that the anomalous Kovacs response
persists in the linear response regime, with its emergence being
controlled by the sign of the second mode of
$\phi_{\theta}$. Therefore, the anomalous Kovacs response is not a
non-linear effect. In fact, the separation between normal and
anomalous behaviour is similar to the one found in the non-linear
regime~\cite{prados_kovacs-like_2014,trizac_memory_2014}: normal
behaviour is found for small inelasticity ($\alpha>\alpha_{c}$),
whereas anomalous behaviour comes about for large inelasticity
($\alpha<\alpha_{c}$).

The linear response approximation makes it possible to analyse the
behaviour of the system for $\uptau\leq\uptau_{w}$. Therein, the
system is relaxing towards the NESS corresponding to the lowest
temperature $T_{1}=T_{\st}-\delta T'$ and the excess kurtosis evolves
according to
\begin{equation}
  \delta A_{2}(\uptau)=\delta\theta_{1}(0)
  \frac{a_{+}a_{-}(\lambda_{+}-\lambda_{-})}
  {M_{12}}
  \left(e^{\lambda_{+}\uptau_{1}}-e^{\lambda_{-}\uptau_{1}}\right).
\end{equation}
This follows from Eq.~\eqref{eq:phi1-2} with the changes
brought about by the different steady temperature:  the initial
temperature jump must be measured with respect to $T_{1}$,
i.e. $\delta\theta_{1}(0)=(T_{0}-T_{1})/T_{1}$, and
$\tau_{1}$ is the scaled time in Eq.~\eqref{eq:scaled-variables-def}
with the substitution $T_{\st}\to T_{1}$. Taking into account
Eq.~\eqref{eq:deltatheta-deltatheta'} and keeping only linear terms in
the deviations,
\begin{equation}\label{eq:A2-kovacs-waiting-window}
  \delta A_{2}(\uptau)=\left(\delta\theta+\delta\theta'\right)
  \frac{a_{+}a_{-}(\lambda_{+}-\lambda_{-})}
  {M_{12}}
  \left(e^{\lambda_{+}\uptau}-e^{\lambda_{-}\uptau}\right).
\end{equation}
Let us recall that $\delta A_{2}(\uptau)>0$ for all times.

\begin{figure}
	\centering
	\includegraphics[width=3.25in]{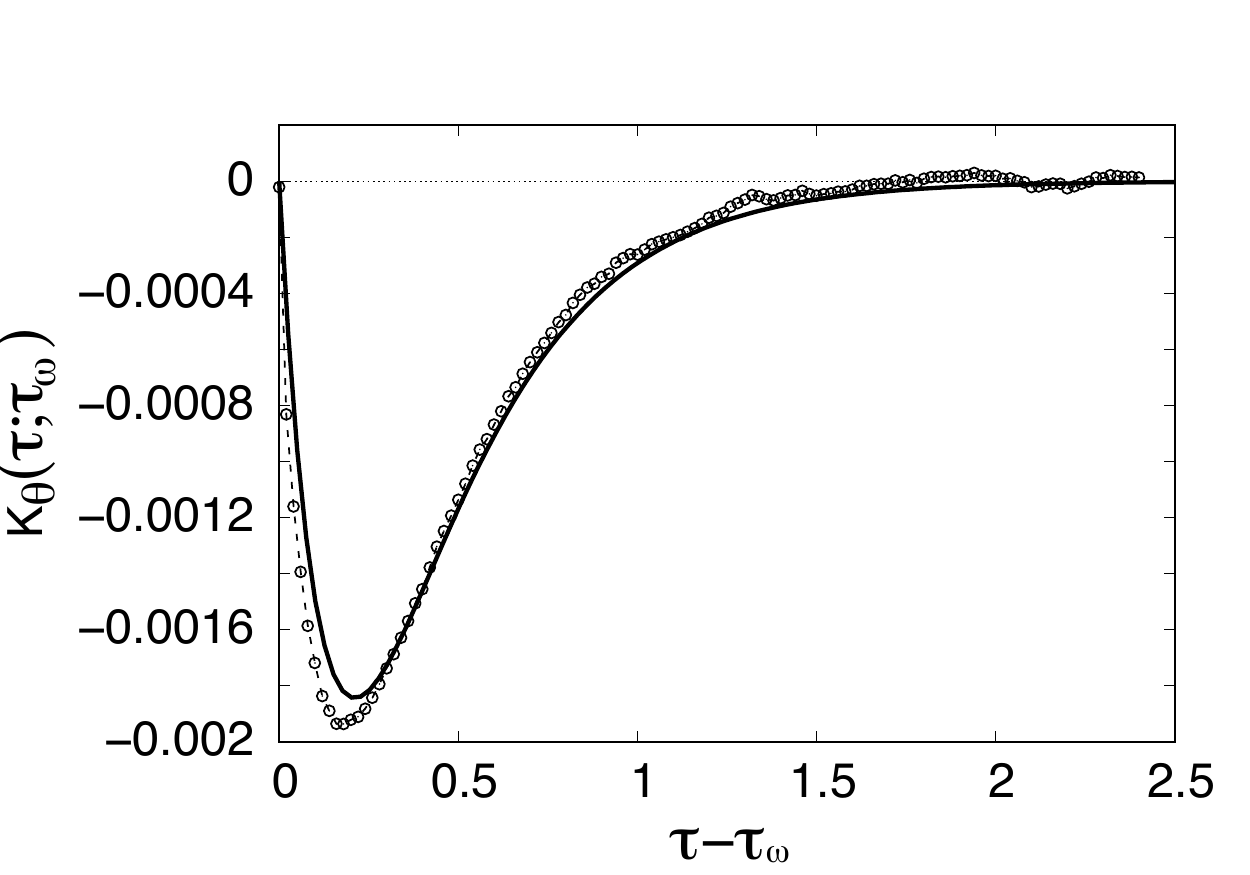}
	\caption{Kovacs hump for large
          inelasticity. Specifically, the restitution coefficient is
          $\alpha=0.3$. The circles are simulation results obtained
          with $N=10^6$ particles averaged over $10^4$ runs, while the
          solid line is our theoretical prediction
          \eqref{eq:phi2-1}. The Kovacs response is anomalous for this
          large inelasticity, $\alpha<\alpha_{c}$.}
	\label{fig:kovacs1}
\end{figure}
\begin{figure}
	\centering
	\includegraphics[width=3.25in]{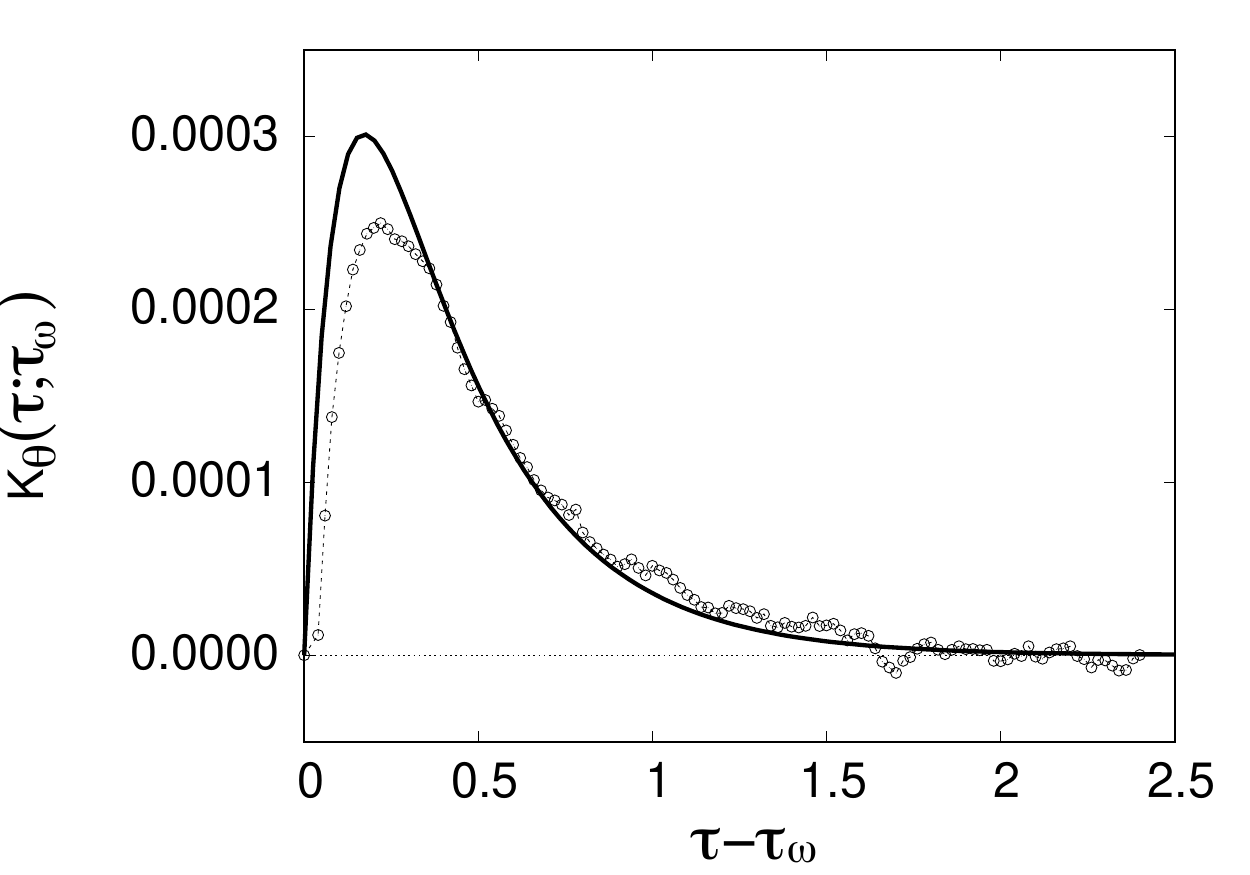}
	\caption{Kovacs hump for small inelasticity.  Specifically,
          the restitution coefficient is $\alpha=0.8$. The response is
          normal in this case, $\alpha>\alpha_{c}$. The circles are
          simulation results obtained with $N=10^6$ particles averaged
          over $5\times 10^4$ runs, whereas the solid line is again
          Eq.~\eqref{eq:phi2-1}.}
	\label{fig:kovacs2}
\end{figure}
In what follows, we compare simulation results from the DSMC numerical
integration of the Boltzmann-Fokker-Planck equation with our
theoretical predictions. Specifically, we plot
\begin{align}
  K_{\theta}(\uptau;\uptau_{w})&\equiv
\frac{\delta\theta(\uptau)}{\delta\! A_{2}(\uptau_{w})}\nonumber\\ &=
\frac{M_{12}}{\lambda_{+}-\lambda_{-}}
\left(e^{\lambda_{+}(\uptau-\uptau_{w})}-e^{\lambda_{-}(\uptau-\uptau_{w})}
\right),\label{eq:phi2-1}
\end{align}
where we have made use of Eqs.~\eqref{eq:Kovacs-delta-theta-tau} and
\eqref{eq:A2-kovacs-waiting-window}, the latter particularised for
$\uptau=\uptau_{w}$. Since $\delta\! A_{2}(\uptau_{w})>0$, we have
that $\delta\theta$ and $K_{\theta}$ have the same sign. In order to
implement the Kovacs protocol we have set the following values for the
noise intensity: $\xi^2=0.2$, $(\xi+\delta \xi)^2=0.35$, and
$(\xi-\delta \xi)^2=0.05$. Fluctuations make these larger jumps
necessary to numerically observe the time evolution of the excess
kurtosis, which is the quantity bringing about the Kovacs effect. As
discussed in Sec.~\ref{sec:linear-relax} and illustrated in
Figs.~\ref{fig:relax-A2} and \ref{fig:relax-T-larger-jumps}, linear
response still holds in this situation~\footnote{This can be
  physically understood in this two-jump experiment, the size of the
  memory effect is quite small.}.

The case $\alpha < \alpha_{c}$, i.e. the large inelasticity regime, is
shown in Fig.~\ref{fig:kovacs1}. Consistently with our theoretical
predictions, the Kovacs response is anomalous, $K_{\theta}<0$, and
the agreement between simulations and theory is excellent.  The small
inelasticity regime, $\alpha>\alpha_{c}$ is illustrated in
Fig. \ref{fig:kovacs2}. Here, the response is normal,
$K_{\theta}>0$, and  the amplitude of the hump is roughly one order of
magnitude smaller than that in Fig. \ref{fig:kovacs1}.

\section{Discussion}\label{sec:discussion}

We have investigated the relaxation of the granular temperature $T$
and the excess kurtosis $a_{2}$---or, alternatively, of the
fourth-moment of the velocity---in the linear response regime. This
study has been carried out by employing two different methods. First,
in the \textit{direct} route, we have linearised the evolution
equations for small changes of the driving and
obtained the analytical solution thereof. Second, in the FDR route, we
have derived a generalised FDR, which relates the relaxation functions
after a small change in the driving with certain time correlation
functions in the NESS.

The theoretical predictions above have been checked against DSMC
results. In the simulations, we have considered both the direct and
the FDR routes. In the first, we have found a perfect agreement
between the simulations and the theoretical predictions. In the
second, the agreement is also very good when the normalised relaxation
function---equal to unity for the initial time---is compared with the
corresponding normalised time correlation in the NESS. However, the
initial value of the time correlation does not match that of the
relaxation function, because two-body correlations have a
non-vanishing contribution. This is analogous to the situation found
in the study of the fluctuations of the total energy of the uniformly
heated granular gas~\cite{garcia_de_soria_energy_2009}.

The linear relaxation function of the granular temperature in the
simulations is almost perfectly fitted by a single exponential, over
the whole range of inelasticities. Interestingly, this seems to rule
out the possibility of memory effects, since it is well-known that
non-exponential relaxation is a prerequisite for the appearance of
aging and memory effects---a simple example is the one-dimensional
Ising
model~\cite{spohn_stretched_1989,brey_stretched_1993,brey_low-temperature_1996,ruiz-garcia_kovacs_2014}. Nevertheless,
the relaxation function is not exactly exponential: the relaxation has
two modes, but the coefficient of the second mode is much smaller than
that of the first mode---their ratio varying from $0$ for $\alpha=1$
(the elastic limit) to roughly $0.01$ for $\alpha=0$ (the completely
inelastic case). It is this smallness that makes inferring the
deviation from the exponential behaviour by looking only at the
numerical data problematic. This highlights the difficulty of
ruling out the possible emergence of memory effects by investigating
only simple, single-jump, relaxation experiments.

Despite its smallness, the non-trivial behaviour of the coefficient of
the second mode as a function of the inelasticity and, in particular,
its changing sign at the critical value $\alpha_{c}$, as shown in
Fig.~\ref{fig:a+-and-a-}, have important physical consequences. The
most striking one is its bringing about the anomalous Kovacs effect in
linear response. In molecular systems, there is a clear parallelism
between the observed universal properties of the Kovacs hump in
experiments---which are done in the non-linear regime---and the
general properties that can be rigorously proved in linear response.
In particular, its normal character---a well-defined sign that does
not change with the system parameters---stem from the equilibrium FDR
that ensures that the coefficients of all the modes in the direct
relaxation function are positive~\cite{prados_kovacs_2010}.

Extension of the above parallelism between the ``empirical''
non-linear results and the theoretical linear response results to
athermal systems---like granular fluids or granular matter---suggests
that it is precisely the change of sign of the coefficient of the
second mode that gives rise in general to the anomalous Kovacs effect,
not only for the granular gas that we have considered here. This
improves our understanding of the emergence of the anomalous Kovacs
effect in athermal systems. Moreover, we have neatly shown that the
anomalous behaviour survives in the linear regime, non-linearities are
not necessary to bring it about.

A perspective for future work is the resolution of the discrepancy
between the initial values of the relaxation function and the
corresponding time correlation function that stems from the
generalised FDR. To accomplish this goal, it is necessary to go beyond
the completely factorised form \eqref{eq:propag-chaos} of the
$N$-particle distribution function, including at least two-body
correlations---but not only in the NESS, as was done in
Ref.~\cite{garcia_de_soria_energy_2009}, but also for a time-dependent
situation.  Another possible avenue for future development is the
analysis of linear response results in other, more complex,
intrinsically non-equilibrium systems, like the rough granular
gas~\cite{kremer_transport_2014,reyes_role_2014,torrente_large_2019}
or active
matter~\cite{baskaran_enhanced_2008,baskaran_nonequilibrium_2010,ihle_kinetic_2011,marchetti_hydrodynamics_2013,ihle_chapmanenskog_2016,bonilla_contrarian_2019}.

\bibliography{Mi-biblioteca-07-jun-2021} 

\end{document}